\documentclass[a4paper,11pt]{article}
\pdfoutput=1 

\usepackage{jheppub} 

\usepackage[T1]{fontenc} 
\usepackage{ulem,slashed}
\usepackage{amsmath,amssymb,epsfig,subfigure,color,soul,fleqn,rotating}

\newcommand{\AddrHEPHY}{%
 \it Institut f\"ur Hochenergiephysik der \"Osterreichischen Akademie
der Wissenschaften, A-1050 Vienna, Austria\\}

\newcommand{\AddrVienna}{
\it Universit\"at Wien, Fakult\"at f\"ur Physik,
A-1090 Vienna, Austria \\}

\newcommand{\AddrGAKUGEI}{%
 \it Department of Physics, Tokyo Gakugei University, Koganei,
Tokyo 184-8501, Japan\\}

\newcommand{\be}{\begin{eqnarray}}
\newcommand{\ee}{\end{eqnarray}}
\newcommand{\nee}{\nonumber\end{eqnarray}}
\newcommand{\nn}{\nonumber\\}
\newcommand{\msbar}{{\overline{\rm MS}}}
\newcommand{\drbar}{{\overline{\rm DR}}}
\newcommand{\thw}{\theta_{\textit{\tiny{W}}}}

\newcommand{\mch}[1] {m_{\ti \x^+_{#1}}}
\newcommand{\mnt}[1] {m_{\ti \x^0_{#1}}}

\newcommand{\msg}    {m_{\ti g}}
\newcommand{\msu}[1] {m_{\ti u_{#1}}}
\newcommand{\msd}[1] {m_{\ti d_{#1}}}

\def\gev             {{\rm GeV}}

\def\be            {\begin{equation}}
\def\ee            {\end{equation}}
\def\bea            {\begin{eqnarray}}
\def\eea            {\end{eqnarray}}

\def\a              {\alpha}
\def\b               {\beta}
\def\d               {\delta}

\def\x               {\chi}
\def\ti              {\tilde}
\def\sq              {\ti q}
\def\st              {\ti t}
\def\sb              {\ti b}

\def\chp             {\ti \x^+}

\def\sg              {\ti g}

\def\su                {\ti{u}}

\def \sca                 {\ti{c}}
\def\sd                {\ti{d}}
\def\ss                  {\ti{s}}

\def\dll            {\d^{LL}_{23}}
\def\durr            {\d^{uRR}_{23}}
\def\durl            {\d^{uRL}_{23}}
\def\dulr            {\d^{uLR}_{23}}

\newcommand{\XX}{{(X^2)}}

\title{\boldmath The decays $h^0 \to b \bar{b}$ and $h^0 \to c \bar{c}$ in the light of the MSSM with quark flavour violation}

\author[a]{H. Eberl}
\author[a,1]{E. Ginina\note{On leave of absence from the Institute of Nuclear Research and Nuclear Energy, Sofia.}}
\author[b]{A. Bartl}
\author[c]{K. Hidaka}
\author[a]{W. Majerotto}

\affiliation[a]{\AddrHEPHY}
\affiliation[b]{\AddrVienna}
\affiliation[c]{\AddrGAKUGEI}

\emailAdd{helmut.eberl@oeaw.ac.at}
\emailAdd{elena.ginina@oeaw.ac.at}
\emailAdd{alfred.bartl@univie.ac.at}
\emailAdd{hidaka@u-gakugei.ac.jp}
\emailAdd{walter.majerotto@oeaw.ac.at}

\abstract{We calculate the decay width of $h^0 \to b \bar{b}$ in the Minimal Supersymmetric Standard Model (MSSM) with quark flavour violation (QFV)
at full one-loop level. We study the effect of $\sca-\st$ mixing and $\ss-\sb$ mixing taking into account the constraints from the B meson data. We 
discuss and compare in detail the decays $h^0 \to c \bar{c}$ and $h^0 \to b \bar{b}$ within the framework of the perturbative mass insertion technique using the Flavour Expansion Theorem.
The deviation of both decay widths from the Standard Model values can be quite large.
Whereas in  $h^0 \to c \bar{c}$ it is almost entirely due to the flavour violating part of the MSSM, in $h^0 \to b \bar{b}$ it is mainly due to the flavour conserving 
part. 
Nevertheless, the QFV contribution to  $\Gamma(h^0 \to b \bar{b})$ due to $\sca-\st$ 
mixing and chargino exchange can go up to $\sim 7\%$.}

\begin{document} 
\maketitle
\flushbottom

\section{Introduction}

In the Standard Model (SM) the Higgs mechanism is responsible for the mass of the fermions. Therefore, it is necessary to measure the 
Yukawa couplings very precisely. Since the Yukawa coupling is proportional to the fermion mass, the largest decay branching ratio
of the Higgs boson, discovered by CMS and ATLAS at LHC~\cite{Aad:2012tfa, Chatrchyan:2012xdj} with a mass of approximately 125 GeV, is that of $h^0 \to b \bar{b}$.
Within the SM this branching ratio is B$(h^0 \to b \bar{b})=0.577~^{\tiny{+3.2\%}}_{\tiny{-3.3\%}}$~\cite{pdg2014}.
Although the Higgs boson properties measured so far are consistent with the SM, deviations from the SM are not yet excluded and could point to "New Physics".

An important extension of the SM is provided by Supersymmetry (SUSY), in particular by the Minimal Supersymmetric Standard Model (MSSM).
In the MSSM, the discovered Higgs boson could be the lightest neutral Higgs boson $h^0$. Quark flavour conservation (QFC) is usually assumed 
(apart from the quark flavour violation (QFV) induced by the Cabibbo-Kobayashi-Maskawa (CKM) matrix). However, SUSY QFV terms could be present in the mass mixing matrix of the squarks, especially mixing terms between the 2nd and the 3rd squark generations. 

In a previous paper~\cite{Bartl:2014bka} we studied the impact of $\sca_{L,R}-\st_{L,R}$ mixing on the decay $h^0 \to c \bar{c}$. We showed that
the deviation from the SM width $\Gamma (h^0 \to c \bar{c})$ can go up to $\pm 35\%$, due to QFV effects at one-loop level. In the 
present paper, we study the influence of this mixing in the decay $h^0 \to b \bar{b}$. (For completeness we have also studied $\ti{s}_{L,R}-\ti{b}_{L,R}$ mixing effects, but they have turned out to be very small.)
There are, however, constraints on the mixing between the $2^{\rm nd}$ and the $3^{\rm rd}$ generations of squarks from B-physics measurements ($\Delta M_{B_s},~ {\rm B}(b \to s \gamma), ~{\rm B}(b \to s l^+ l^-),
~{\rm B}(B_s \to \mu^+ \mu^-),~ {\rm B}(B^+ \to \tau^+ \nu)$), as well as from $m_{h^0}$ measurements and SUSY particle searches. We take into account all these constraints.

First, in our calculation of $\Gamma (h^0 \to b \bar{b})$ at full one-loop level, we will largely proceed analogously to the case of $h^0 \to c \bar{c}$~\cite{Bartl:2014bka, Eberl:2014dla, Hidaka:2015zqa,Ginina:2015doa, Hidaka:2015ewt}, except for the particular features characteristic of the decays into bottom quarks, as the large $\tan \b$ enhancement and resummation of the bottom Yukawa coupling.

The main new feature in this paper is the additional adoption of the perturbative mass insertion technique using the Flavour Expansion Theorem~\cite{Dedes:2015twa}. We will discuss it both in the  $h^0 \to c \bar{c}$ and $h^0 \to b \bar{b}$ case. It gives systematic insight into the various QFV contributions. 
In particular, we show that due to the fact that the product $T^U_{32} M^U_{23}$ is {\it apriori} unbounded by experiment, the correction to the width of  $h^0 \to c \bar{c}$ can become large so that perturbation theory breaks down.
(For the definitions of  $T^U$ and $M^U$ see eqs.~(\ref{EqMassMatrix1}), (\ref{EqM2LLRR}) and (\ref{EqMassMatrix}) below.) In the $h^0 \to b \bar{b}$ case this is not possible.

%
%
\section{Definition of the QFV parameters}
\label{sec:sq.matrix}

In the MSSM's super-CKM basis of $\sq_{0 \gamma} =
(\sq_{1 {\rm L}}, \sq_{2 {\rm L}}, \sq_{3 {\rm L}}$,
$\sq_{1 {\rm R}}, \sq_{2 {\rm R}}, \sq_{3 {\rm R}}),~\gamma = 1,...6,$  
with $(q_1, q_2, q_3)=(u, c, t),$ $(d, s, b)$, one can write the squark mass matrices in their most general $3\times3$-block form~\cite{Allanach:2008qq}
\begin{equation}
    {\cal M}^2_{\tilde{q}} = \left( \begin{array}{cc}
        {\cal M}^2_{\tilde{q},LL} & {\cal M}^2_{\tilde{q},LR} \\[2mm]
        {\cal M}^2_{\tilde{q},RL} & {\cal M}^2_{\tilde{q},RR} \end{array} \right),
 \label{EqMassMatrix1}
\end{equation}
with $\tilde{q}=\tilde{u},\tilde{d}$. The left-left and right-right blocks in eq.~(\ref{EqMassMatrix1}) are given by
\begin{eqnarray}
    & &{\cal M}^2_{\tilde{u},LL} = V_{\rm CKM} M_Q^2 V_{\rm CKM}^{\dag} + D_{\tilde{u},LL}{\bf 1} + \hat{m}^2_u, \nonumber \\
    & &{\cal M}^2_{\tilde{u},RR} = M_U^2 + D_{\tilde{u},RR}{\bf 1} + \hat{m}^2_u, \nonumber \\
    & & {\cal M}^2_{\tilde{d},LL} = M_Q^2 + D_{\tilde{d},LL}{\bf 1} + \hat{m}^2_d,  \nonumber \\
    & & {\cal M}^2_{\tilde{d},RR} = M_D^2 + D_{\tilde{d},RR}{\bf 1} + \hat{m}^2_d,
     \label{EqM2LLRR}
\end{eqnarray}
where $M_{Q,U,D}$ are the hermitian soft SUSY-breaking mass matrices of the squarks and
$\hat{m}_{u,d}$ are the diagonal mass matrices of the up-type and down-type quarks.
Furthermore, 
$D_{\tilde{q},LL} = \cos 2\beta m_Z^2 (T_3^q-e_q
\sin^2\theta_W)$ and $D_{\tilde{q},RR} = e_q \sin^2\theta_W \times$ $ \cos 2\beta m_Z^2$,
where
$T_3^q$ and $e_q$ are the isospin and
electric charge of the quarks (squarks), respectively, and $\theta_W$ is the weak mixing
angle.
Due to the $SU(2)_{\rm L}$ symmetry the left-left blocks of the up-type and down-type squarks in eq.~(\ref{EqM2LLRR}) are related
by the CKM matrix $V_{\rm CKM}$.
The left-right and right-left blocks of eq.~(\ref{EqMassMatrix1}) are given by
\begin{eqnarray}
 {\cal M}^2_{\tilde{u},RL} = {\cal M}^{2\dag}_{\tilde{u},LR} &=&
\frac{v_2}{\sqrt{2}} T_U - \mu^* \hat{m}_u\cot\beta, \nonumber \\
 {\cal M}^2_{\tilde{d},RL} = {\cal M}^{2\dag}_{\tilde{d},LR} &=&
\frac{v_1}{\sqrt{2}} T_D - \mu^* \hat{m}_d\tan\beta,
\label{M2sqdef}
\end{eqnarray}
where $T_{U,D}$ are the soft SUSY-breaking trilinear 
coupling matrices of the up-type and down-type squarks entering the Lagrangian 
${\cal L}_{int} \supset -(T_{U\alpha \beta} \ti{u}^\dagger _{R\a}\ti{u}_{L\b}H^0_2 $ 
$+ T_{D\alpha \beta} \ti{d}^\dagger _{R\a}\ti{d}_{L\b}H^0_1)$,
$\mu$ is the higgsino mass parameter, and $\tan\beta$ is the ratio of the vacuum expectation values of the neutral Higgs fields $v_2/v_1$, with $v_{1,2}=\sqrt{2} \left\langle H^0_{1,2} \right\rangle$.
The squark mass matrices are diagonalized by the $6\times6$ unitary matrices $U^{\tilde{q}}$,
$\tilde{q}=\tilde{u},\tilde{d}$, such that
\begin{eqnarray}
&&U^{\tilde{q}} {\cal M}^2_{\tilde{q}} (U^{\tilde{q} })^{\dag} = {\rm diag}(m_{\tilde{q}_1}^2,\dots,m_{\tilde{q}_6}^2)\,,
\label{Umatr}
\end{eqnarray}
with $m_{\tilde{q}_1} < \dots < m_{\tilde{q}_6}$.
The physical mass eigenstates
$\sq_i, i=1,...,6$ are given by $\sq_i =  U^{\sq}_{i \alpha} \sq_{0\alpha} $.

We define the QFV parameters in the up-type squark sector 
$\delta^{LL}_{\alpha\beta}$, $\delta^{uRR}_{\alpha\beta}$
and $\delta^{uRL}_{\alpha\beta}$ $(\alpha \neq \beta)$ as follows \cite{Gabbiani:1996hi}:
\begin{eqnarray}
\delta^{LL}_{\alpha\beta} & \equiv & M^2_{Q \alpha\beta} / \sqrt{M^2_{Q \alpha\alpha} M^2_{Q \beta\beta}}~,
\label{eq:InsLL}\\[3mm]
\delta^{uRR}_{\alpha\beta} &\equiv& M^2_{U \alpha\beta} / \sqrt{M^2_{U \alpha\alpha} M^2_{U \beta\beta}}~,
\label{eq:InsRR}\\[3mm]
\delta^{uRL}_{\alpha\beta} &\equiv& (v_2/\sqrt{2} ) T_{U\alpha \beta} / \sqrt{M^2_{U \alpha\alpha} M^2_{Q \beta\beta}}~,
\label{eq:InsRL}
\end{eqnarray}
where $\alpha,\beta=1,2,3 ~(\alpha \ne \beta)$ denote the quark flavours $u,c,t$. Analogously, for the down-type squark sector we have
\begin{eqnarray}
\delta^{dRR}_{\alpha\beta} &\equiv& M^2_{D \alpha\beta} / \sqrt{M^2_{D \alpha\alpha} M^2_{D \beta\beta}}~,
\label{eq:dRR}\\[3mm]
\delta^{dRL}_{\alpha\beta} &\equiv& (v_1/\sqrt{2} ) T_{D\alpha \beta} / \sqrt{M^2_{D \alpha\alpha} M^2_{Q \beta\beta}}~,
\label{eq:dRL}
\end{eqnarray}
and the parameter $\delta^{LL}_{\alpha\beta}$ is defined by eq.(\ref{eq:InsLL}). The subscripts $\alpha,\beta=1,2,3 ~(\alpha \ne \beta)$ denote the quark flavours $d,s,b$.

In this paper we focus on the $\ti{c}_R - \ti{t}_L$, $\ti{c}_L - \ti{t}_R$, $\ti{c}_R - \ti{t}_R$, and $\ti{c}_L - \ti{t}_L$ mixing 
which is described by the QFV parameters $\delta^{uRL}_{23}$, 
$\delta^{uLR}_{23} \equiv ( \delta^{uRL}_{32})^*$, $\delta^{uRR}_{23}$, and $\dll$, respectively. The
$\ti{t}_R - \ti{t}_L$ mixing is described by the QFC parameter $\delta^{uRL}_{33}$.
We also allow $\ss - \sb$ mixing. All parameters are assumed to be real, i.e. no CP-violation is considered. In principle, there might be in addition also trilinear non-holomorphic interactions, see eq.~(1.5) in~\cite{Dedes:2014asa}. These interactions are not taken into account in this study.

\section{$h^0 \to b \bar{b}$ at one-loop level with flavour violation}
\label{sec:h2bbb}
%

We write the decay width of $h^0 \to b \bar{b}$ including the one-loop contributions as
\be
\Gamma(h^0 \to b \bar{b})=\Gamma^{\rm tree}(h^0 \to b \bar{b})+\delta \Gamma^{\rm  1loop}(h^0 \to b \bar{b})
\label{decaywidth}
\ee
with the tree-level decay width
\be
\Gamma^{\rm tree}(h^0 \to b \bar{b})=\frac{\rm N_C}{8 \pi} m_{h^0} (s_1^b)^2 \bigg( 1- \frac{4 m_b^2}{m_{h^0}^2}\bigg)^{3/2}\,,
\label{decaywidttree}
\ee
where ${N_C}=3$, $m_{h^0}$ is the on-shell mass of $h^0$ and the tree-level coupling $s_1^b$ is
\be
 s_1^b=g \frac{m_b}{2 m_W} \frac{\sin{\a}}{\cos{\b}} =\frac{h_b}{\sqrt{2}}\sin{\a}\,,
 \label{treecoup}
\ee
$\alpha$ is the mixing angle of the two CP-even Higgs bosons, $h^0$ and $H^0$~\cite{G&H}.

In the calculation of $\delta \Gamma^{\rm  1loop}(h^0 \to b \bar{b})$ we proceed in a way analogously to the calculation of $\delta \Gamma^{\rm  1loop}(h^0 \to c \bar{c})$ in Ref.~\cite{Bartl:2014bka}.
In addition to the diagrams that contribute within the SM, $\delta \Gamma^{\rm  1loop}(h^0 \to b \bar{b})$ receives contributions from the exchange of SUSY particles and Higgs bosons. The corresponding diagrams are shown in Fig.~2 of \cite{Bartl:2014bka}, replacing $c$ by $b$ quarks and $\su \leftrightarrow \sd$. The dominant SUSY contribution is due to gluino and chargino exchange. The gluino and the chargino contribute also to the self-energy of the b quark.

As in Ref.~\cite{Bartl:2014bka} we use the $\drbar$ renormalisation scheme, where all input parameters in the Lagrangian (masses, fields and coupling parameters) 
are UV finite, defined at the scale $Q=1~\rm TeV$. In order to obtain the shifts from the $\drbar$ masses and fields to the physical scale-independent quantities, we use on-shell renormalisation conditions. Moreover, we include in our calculations the contributions from real hard gluon/photon radiation from the final b quarks.

The one-loop corrected width $\Gamma(h^0 \to b \bar{b})$ is therefore given by 
\be
\Gamma(h^0 \to b \bar{b}) = \Gamma^{g, \rm impr} + \d \Gamma ^{\sg} + \d \Gamma ^{EW}\,,
\label{eqGamma}
\ee
where $ \Gamma^{g, \rm impr}$ includes the tree-level and the gluon loop contribution, see eq.(55) in~\cite{Bartl:2014bka}, $\d \Gamma ^{\sg}$  is the gluino one-loop contribution and $ \d \Gamma ^{EW}$ is the electroweak one-loop contribution. Moreover, we have considered the large $\tan \beta$ enhancement and the resummation of the bottom Yukawa coupling~\cite{Carena:1999py}. 
It turns out, however, that in our case with large $m_{A^0}$ close to the decoupling limit, the resummation effect is very small ($<0.1\%$).

 %
%
\section{Mass Insertion technique}
\label{Sec:MI}

In this section, we want to apply to the decays $h^0 \to c \bar{c}$ and  $h^0 \to b \bar{b}$ the mass insertion technique as well as the Flavour Expansion Theorem (FET) as developed by Dedes et al. in~\cite{Dedes:2015twa}.
Let us consider the expression
\begin{equation}
X = U^{\tilde q}_{i A} U^{\tilde q *}_{i B} B_0(0, m^2, m^2_{\tilde q_i}) \, ,
\end{equation}
with $A \ne B$. $U^{\sq}$ are defined with eq. (\ref{Umatr}) and $B_0$ are the two-point Passarino-Veltman functions.
$X$ given in terms of mass eigenstates
can be expanded into mass insertions (MIs) by the FET~\cite{Dedes:2015twa} 
\begin{eqnarray}
X & =  & M^I_{A B} \, b_0\! \left(1,  m^2, \{M_{A A}, M_{B B}\}\right) \nn
   & + &   M^I_{A i} M^I_{i B} \, b_0\! \left(2,  m^2, \{M_{A A}, M_{i i},  M_{B B}\} \right) \nn
   & + &   M^I_{A i} M^I_{i j} M^I_{j B} \, b_0\! \left(3,  m^2, \{M_{A A}, M_{i i}, M_{j j}, M_{B B}\} \right) \nn
   & + &   M^I_{A i} M^I_{i j} M^I_{j k}  M^I_{k B} \, b_0\! \left(4,  m^2, \{M_{A A}, M_{i i}, M_{j j}, M_{k k},  M_{B B}\} \right) + \ldots \, ,
\label{mass2MI_exp4}   
\end{eqnarray}
by using Einstein summation convention.
The diagonal elements of the squared mass matrix are denoted by $M_{i i}$, and the off-diagonal ones by the matrix $M^I$ with
the restriction $M^I_{i i} = 0$. This formula and all following MI formulas in this section have been checked with the 
Mathematica package {\tt MassToMI}~\cite{arXiv:1509.05030}.
The generalized $b_0$ functions~\cite{Dedes:2015twa}, where the first argument shows how many insertions are done,
can be written recursively as
\begin{eqnarray}
 b_0(1,a,\{b,c\}) & = &  \frac{b_0(a,b)-b_0(a,c)}{b-c}\,,\nn
 b_0(2,a,\{b,c,d\}) & = &  \frac{b_0(1,a,\{b,c\})  - b_0(1,a,\{b,d\}) }{c-d}\,, \nn
 b_0(3,a,\{b,c,d,e\}) & = &  \frac{b_0(2,a,\{b,c, d\})  - b_0(2,a,\{b,c, e\}) }{d-e}\,, \nn
  b_0(4,a,\{b,c,d,e,f\}) & = &  \frac{b_0(3,a,\{b,c,d,e\})  - b_0(3,a,\{b,c,d,f\}) }{e-f}\,,
\end{eqnarray}
with 
\begin{equation}
b_0(a,b) \equiv  B_0(0, a,b) = \frac{b \log \left(\frac{b}{Q^2}\right)-a \log
   \left(\frac{a}{Q^2}\right)}{a-b}+ \Delta +1\,,
\end{equation}
with the renormalisation scale $Q$ and $\Delta$ denotes the UV-divergence parameter.
These functions are totally symmetric under any permutation of the set of arguments in the curly brackets.
Note that $b_0(1,a,\{b,c\})\equiv c_0(a,b,c)\equiv C_0(0,0,0,a,b,c)$,~$ b_0(2,a,\{b,c,d\}) \equiv D_0(0,0,0,0,0,0,a,b,c,d)$, etc., where 
$C_0$ and $D_0$ are the scalar 3-point and 4-point Passarino-Veltman functions~\cite{PV}.
The general formula for a number of degenerate arguments is useful~\cite{Dedes:2015twa},
\begin{equation}
\lim_{ \{x_0, \ldots, x_m \} \to \{ \xi, \ldots, \xi \} } 
b_0\! \left(k, a, \{ x_0, \ldots, x_k \}\right) = {1 \over m!} { \partial^m \over \partial \xi^m} 
b_0\! \left(k - m, a, \{ \xi, x_{m+1}, \ldots, x_k\} \right)\, ,
\label{b0_degenerate}
\end{equation}
for $k \ge 1$ and $m \le k$.
The derivative of $b_0$ with respect to the second argument reads
\begin{equation}
b_0{}^{(0,1)}(a,b) = \frac{1}{a-b}+\frac{a \log \left(\frac{b}{a}\right)}{(a-b)^2}\,.
\end{equation}
The derivative of $b_0$ with respect to the first argument can be written as
\begin{equation}
b_0{}^{(1,0)}(a,b) =  b_0{}^{(0,1)}(b,a)\, .
\end{equation}
By using eq.~(\ref{b0_degenerate}) we can write $b_0(1,a,\{b,b\})$ as  $b_0{}^{(0,1)}(a,b)$.

\subsection{Gluino contribution to \boldmath $h^0 \to c \bar c$}
\label{SecMI:gluino2ccb}

\begin{figure*}[h!]
\centering
\subfigure[]{
   { \mbox{\hspace*{-1cm} \resizebox{6.cm}{!}{\includegraphics{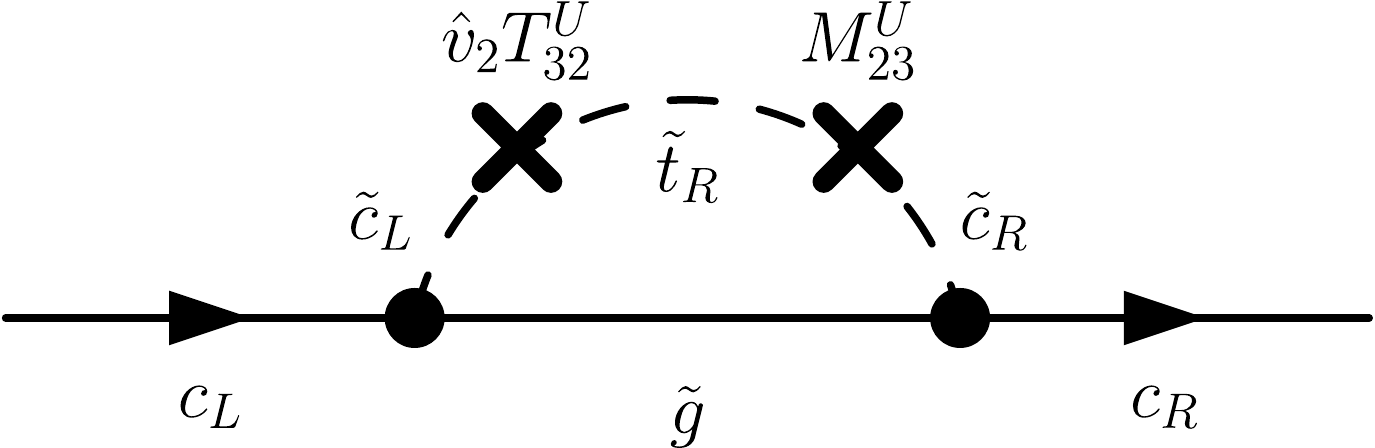}} \hspace*{-0.8cm}}}
   \label{diag_ccb1}} \qquad
 \subfigure[]{
   { \mbox{\hspace*{+0.cm} \resizebox{6.cm}{!}{\includegraphics{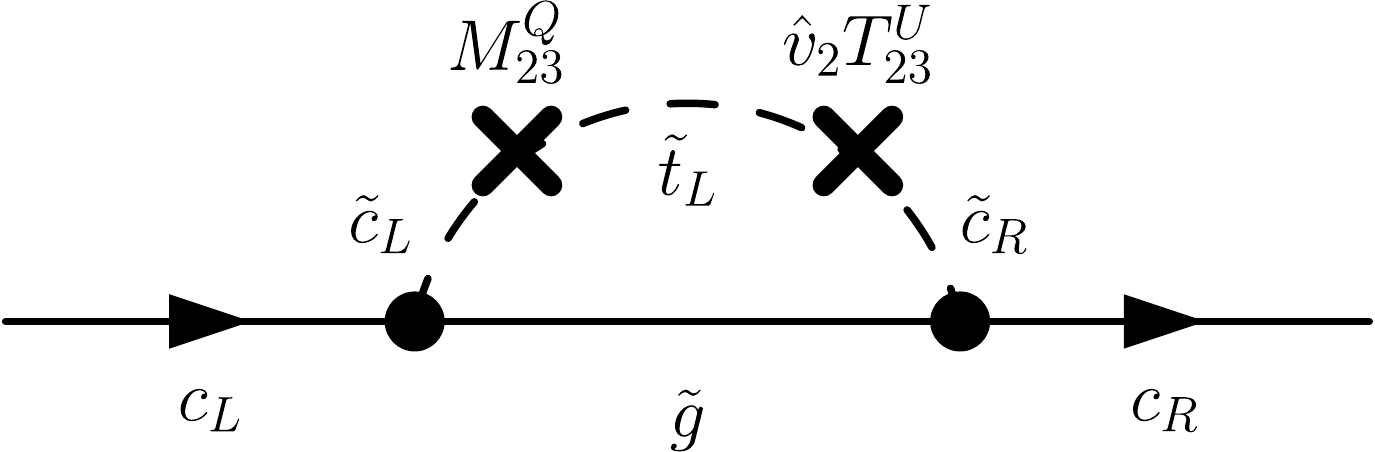}} \hspace*{-1cm}}}
  \label{diag_ccb2}
  }\\
 \subfigure[]{
   { \mbox{\hspace*{-1cm} \resizebox{6.cm}{!}{\includegraphics{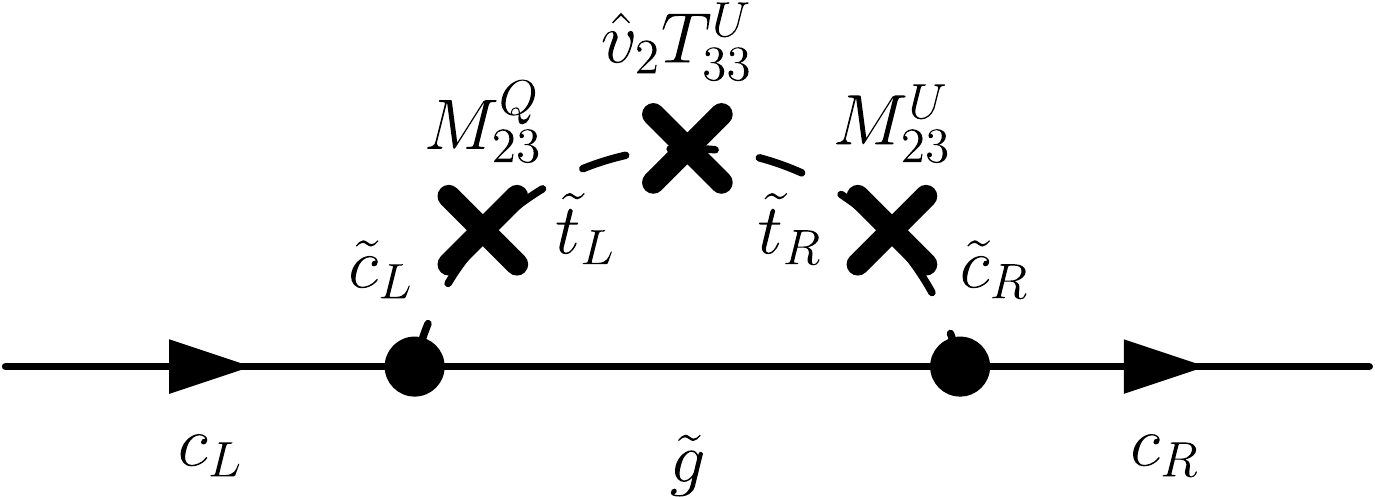}} \hspace*{-0.8cm}}}
   \label{diag_ccb3}} \qquad
 \subfigure[]{
   { \mbox{\hspace*{+0.cm} \resizebox{6.cm}{!}{\includegraphics{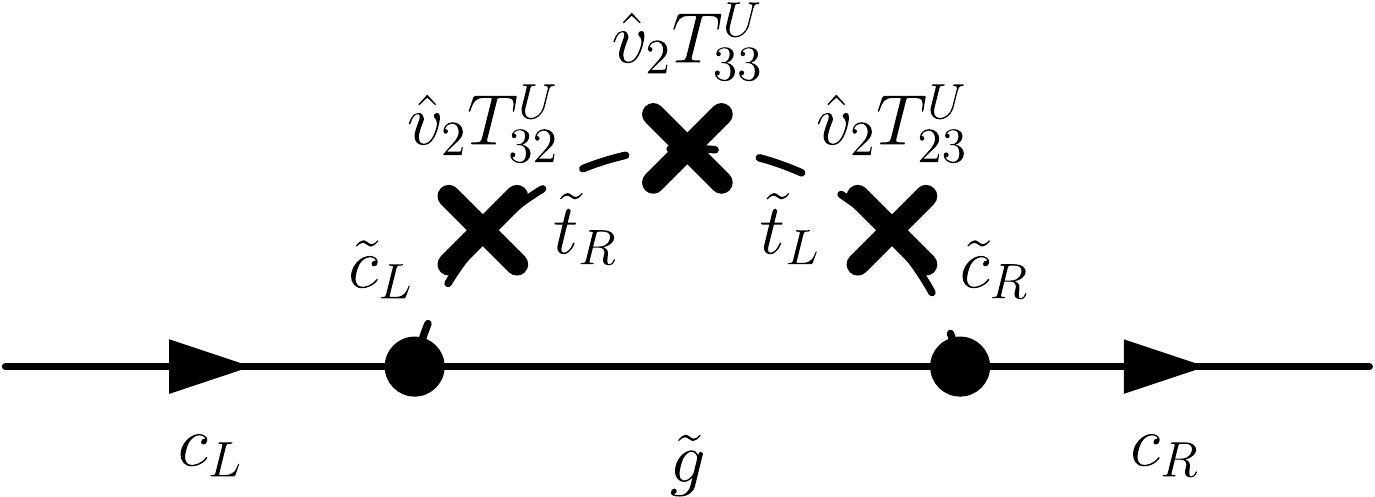}} \hspace*{-1cm}}}
  \label{diag_ccb4}
  } \caption{Quark-flavour violating mass insertions to the charm quark self-energy with gluino, corresponding to
  $T_2$ and $T_3$ in eq.~(\ref{MI_Ts}).}
\label{diag_ccb}
\end{figure*}

As a first example, we want to calculate the self-energy of the c-quark due to $\sg$ and $\su_i$ in the loop.
We decompose the charm self-energy $\Sigma_c$ defined by the Lagrangian
${\cal L} = - \bar c\, \Sigma_c\, c$,
\begin{equation}
\Sigma_c( p ) = \slashed{p} \left( \Sigma^{LL}_c(p^2) P_L + \Sigma^{RR}_c(p^2) P_R \right) +
m_c \left( \Sigma^{RL}_c(p^2) P_L + \Sigma^{LR}_c(p^2) P_R \right)  \, ,
\label{sigma_decomp_c}
\end{equation}
with $\Sigma^{LR}_c = \Sigma^{RL *}_c$. 
We assume real input parameters, therefore $\Sigma^{LR}_c = \Sigma^{RL}_c$, and
\begin{equation}
\Sigma^{LR, \tilde g}_c =   - \frac{2 \a_s}{3 \pi } \frac{\msg}{m_{c}} \sum_{i=1}^6 U^{\su *}_{i2}U^{\su}_{i5}
B_0(m_c^2, m^2_{ \tilde g}, m^2_{\tilde u_i})\, .
\label {dsigLRsg_c}
\end{equation}
Allowing the squared $\tilde u$-mass matrix (eq. (\ref{EqMassMatrix1})) in the form
\begin{equation}
\hspace*{-1cm}    {\cal M}^2_{\tilde{u}} = \left( \begin{array}{cc}
        {\cal M}^2_{\tilde{u},LL} & {\cal M}^2_{\tilde{u},LR} \\[2mm]
        {\cal M}^2_{\tilde{u},RL} & {\cal M}^2_{\tilde{u},RR} \end{array} \right) \equiv M_{ij} =          
         \left( \begin{array}{cccccc}
         M^{LL}_{11} & 0 & 0 & 0 & 0 & 0 \\
         0  & M^{LL}_{22} & M^Q_{23} & 0 & 0 & \hat v_2 T^U_{32} \\
         0 & M^Q_{23} & M^{LL}_{33} & 0 &  \hat v_2 T^U_{23} & \hat v_2 T^U_{33}\\
         0 & 0 & 0 &  M^{RR}_{11} & 0 & 0\\
         0 & 0 &   \hat v_2 T^U_{23} & 0 & M^{RR}_{22} &  M^U_{23} \\
         0 & \hat v_2 T^U_{32} & \hat v_2 T^U_{33} & 0 &  M^U_{23} & M^{RR}_{33}
         \end{array}\right)  \, ,     
 \label{EqMassMatrix}
\end{equation}
with $\hat v_2 = v \sin\beta/\sqrt{2} \sim $ 170~GeV, and the QFV elements of the  $3 \times 3$ matrices $M^2_Q$ and $M^2_U$ are
written as $M^Q_{ij}$ and $M^U_{ij}$, respectively. We neglect the terms proportional to $\mu/\tan\beta$ assuming that $\tan\beta$
is large. The matrix elements $M_{25} = M_{52} = \hat v_2 T^U_{22}$ 
are assumed to be zero, because $T^U_{22}$
is strongly constrained by the colour-breaking condition being proportional to the squared charm-Yukawa coupling (see Appendix D of~\cite{Bartl:2014bka}).
Using eq.~(\ref{mass2MI_exp4}) we get 
\begin{equation}
m_c \Sigma^{LR, \tilde g}_c =   - \frac{2 \a_s}{3 \pi } \msg (T_2 + T_3 + T_4 + \ldots)\,,
\label {dsigLRsg_c1}
\end{equation}
where the QFV contributions read
\begin{eqnarray}
T_2 & = & \hat v_2 T^U_{32} M^U_{23}  \, b_0\! \left(2,  \msg^2, \{ M^{LL}_{22}, M^{RR}_{33}, M^{RR}_{22}\} \right) \nn
&+&          \hat v_2 T^U_{23} M^Q_{23}  \, b_0\! \left(2,  \msg^2, \{M^{LL}_{22}, M^{LL}_{33}, M^{RR}_{22}\} \right) \nn
T_3 & = &  \hat v_2 T^U_{33} \left( M^Q_{23}  M^U_{23}  +  3^\rho\, {\hat v_2}^2  T^U_{23}  T^U_{32} \right)
   \, b_0\! \left(3,  \msg^2, \{ M^{LL}_{22}, M^{LL}_{33}, M^{RR}_{33}, M^{RR}_{22}\} \right)\nn
   T_4 & = &  \hat v_2   T^U_{23} (M^Q_{23})^3  \, b_0\! \left(4,  \msg^2, \{ M^{LL}_{22}, M^{LL}_{33}, M^{LL}_{22},  M^{LL}_{33},  M^{RR}_{22}\} \right)   \nn
          & + &   \hat v_2  T^U_{23} M^Q_{23} (M^U_{23})^2 \, b_0\! \left(4,  \msg^2, \{ M^{LL}_{22}, M^{LL}_{33}, M^{RR}_{22},  M^{RR}_{33},  M^{RR}_{22}\} \right)   \nn  
          & + &   \hat v_2  T^U_{32} (M^U_{23})^3  \, b_0\! \left(4,  \msg^2, \{ M^{LL}_{22}, M^{RR}_{33}, M^{RR}_{22},  M^{RR}_{33},  M^{RR}_{22}\} \right)   \nn    
          & + &   \hat v_2  T^U_{32} M^U_{23} (M^Q_{23})^2 \, b_0\! \left(4,  \msg^2, \{ M^{LL}_{22}, M^{LL}_{33}, M^{LL}_{22},  M^{RR}_{33},  M^{RR}_{22}\} \right)   \nn  
          & + & 3^\rho \Big( {\hat v_2}^3 (T^U_{23})^3 M^Q_{23}  \, b_0\! \left(4,  \msg^2, \{ M^{LL}_{22}, M^{LL}_{33}, M^{RR}_{22},  M^{LL}_{33},  M^{RR}_{22}\} \right)   \nn
          & + & {\hat v_2}^3 (T^U_{23})^2 T^U_{32}  M^U_{23}  \, b_0\! \left(4,  \msg^2, \{ M^{LL}_{22}, M^{RR}_{33}, M^{RR}_{22},  M^{LL}_{33},  M^{RR}_{22}\} \right)   \nn
          & + &  {\hat v_2}^3 (T^U_{32})^2 T^U_{23}  M^Q_{23}  \, b_0\! \left(4,  \msg^2, \{ M^{LL}_{22}, M^{RR}_{33}, M^{LL}_{22},  M^{LL}_{33},  M^{RR}_{22}\} \right)   \nn
          & + &  {\hat v_2}^3 (T^U_{32})^3 M^U_{23}  \, b_0\! \left(4,  \msg^2, \{ M^{LL}_{22}, M^{RR}_{33}, M^{LL}_{22},  M^{RR}_{33},  M^{RR}_{22}\} \right)   \nn
          & + &  {\hat v_2}^3 (T^U_{33})^2 T^U_{23}  M^Q_{23}  \, b_0\! \left(4,  \msg^2, \{ M^{LL}_{22}, M^{LL}_{33}, M^{RR}_{33},  M^{LL}_{33},  M^{RR}_{22}\} \right)   \nn
          & + &  {\hat v_2}^3 (T^U_{33})^2 T^U_{32}  M^U_{23}  \, b_0\! \left(4,  \msg^2, \{ M^{LL}_{22}, M^{RR}_{33}, M^{LL}_{33},  M^{RR}_{33},  M^{RR}_{22}\} \right) \Big)\, , 
\label{MI_Ts}
\end{eqnarray}
with $\rho = 0$. The graphs corresponding to the terms $T_2$ and $T_3$ are given in Figs.~\ref{diag_ccb1} and~\ref{diag_ccb2} or~\ref{diag_ccb3} and~\ref{diag_ccb4}, respectively.
Note that there is no contribution with no mass insertion because we have a helicity flip, and also practically no contribution with 
only one insertion, because $T^U_{22}$ is very small. 
Thus, all terms in eq.~(\ref{MI_Ts}) are quark-flavour violating.
The interactions related to the mass insertions are given by the effective Lagrangian
\bea 
\hspace*{-1cm}{\cal L} = -T^U_{33}\, {\tilde t}_R^{*}  {\tilde t}_L H^0_2 - T^U_{32}\,  {\tilde t}_R^{*}  {\tilde c}_L H^0_2 - T^U_{23}\, {\tilde c}_R^{*} {\tilde t}_L H^0_2 
- M^Q_{23}\, {\tilde t}_L^{*} {\tilde c}_L - M^U_{23}\, {\tilde t}_R^{*} {\tilde c}_R + {\rm h.c.}\,,
\label{}
\eea
with  $H^0_2 = {1 \over \sqrt2}(v \sin\beta +  h^0 \cos\alpha+ \ldots)$.

\begin{figure*}[h!]
\centering
\subfigure[]{ \resizebox{4.5cm}{!}{\includegraphics{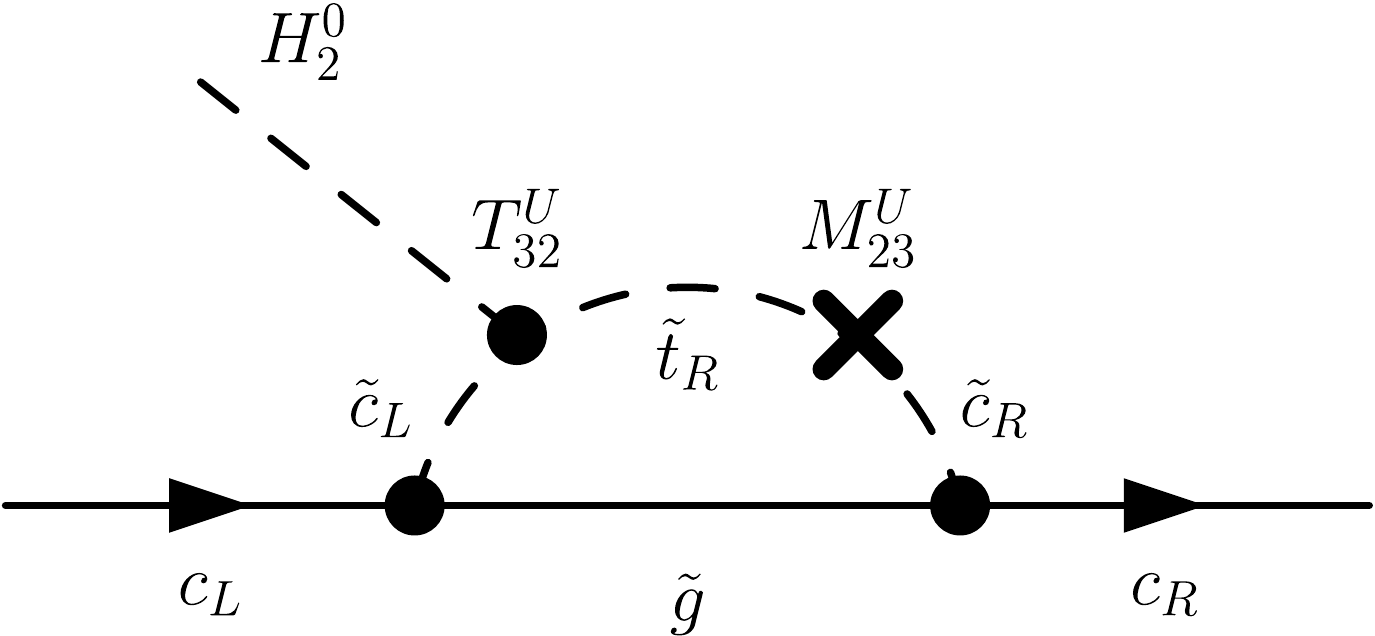}} \label{diag_h0ccb1}} 
\subfigure[]{ \resizebox{4.5cm}{!}{\includegraphics{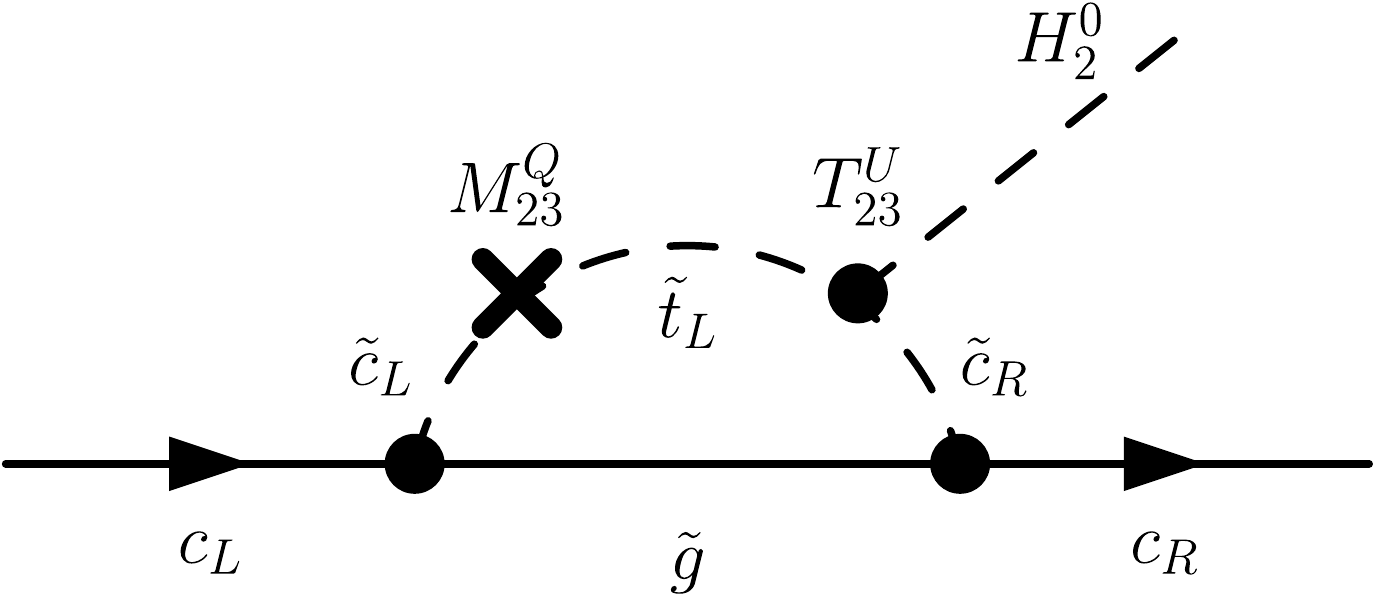}} \label{diag_h0ccb2}} 
\subfigure[]{ \resizebox{4.5cm}{!}{\includegraphics{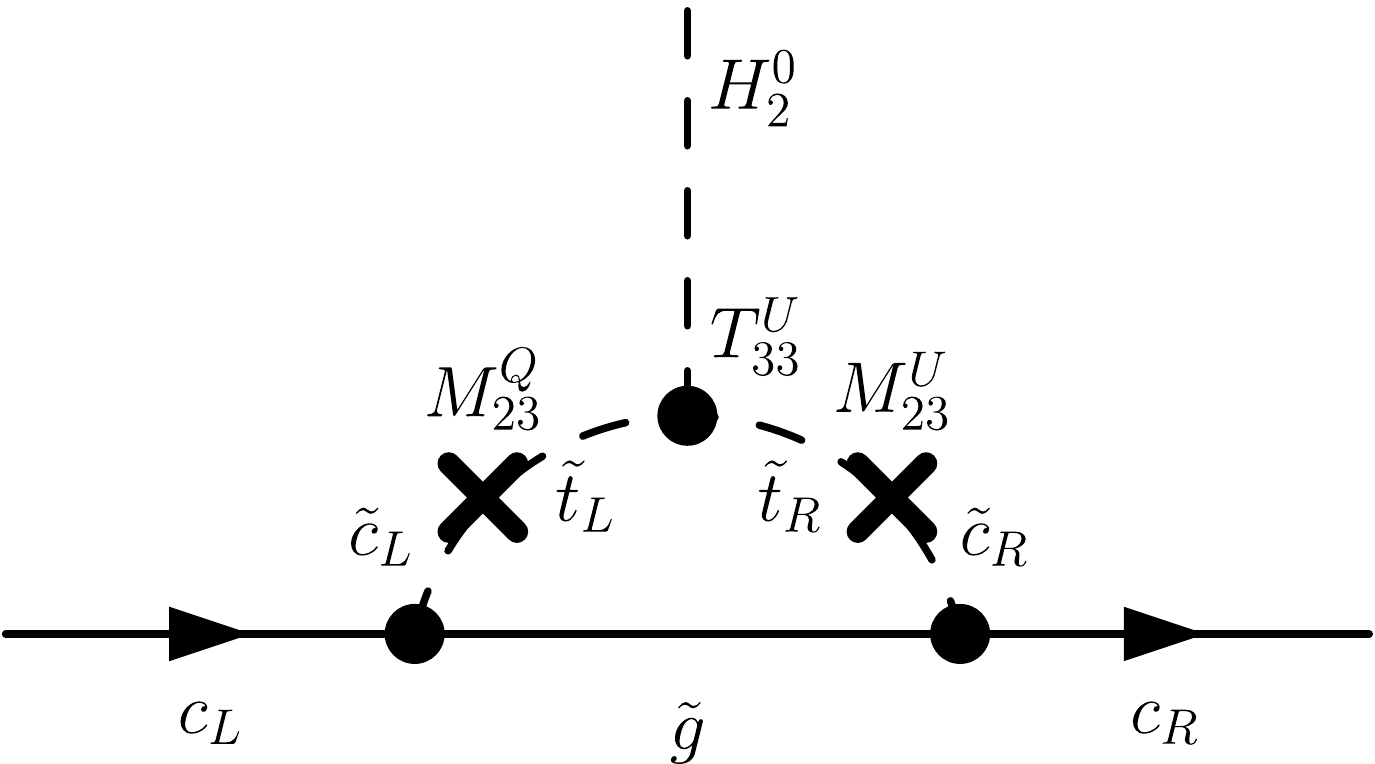}} \label{diag_h0ccb3}} \\
\subfigure[]{ \resizebox{4.5cm}{!}{\includegraphics{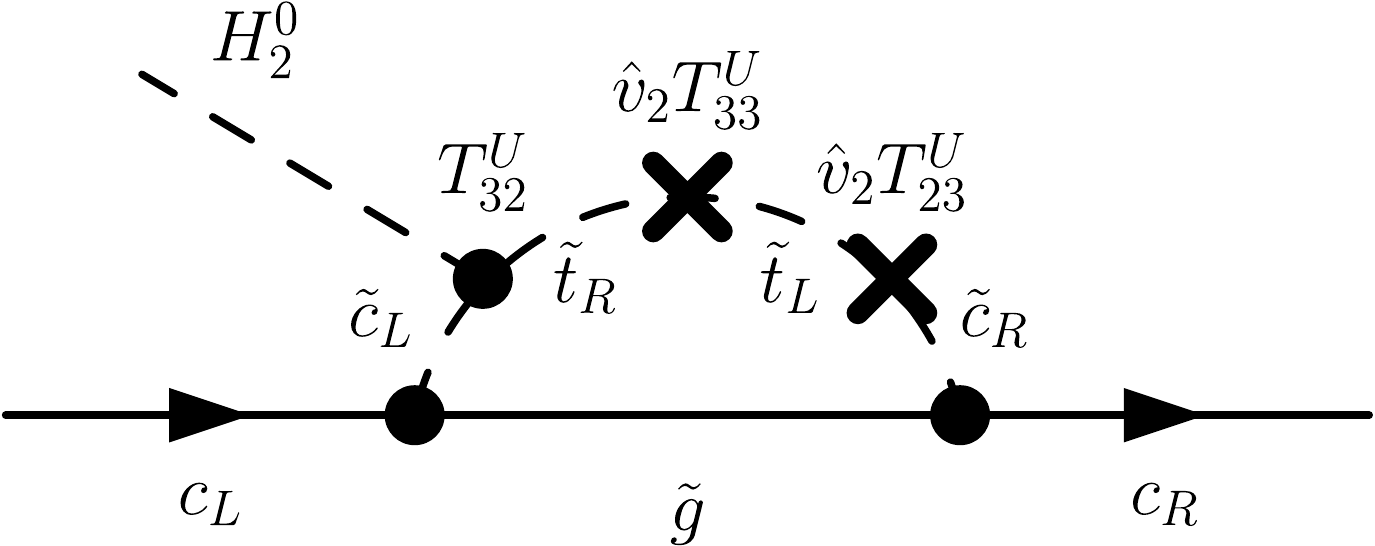}} \label{diag_h0ccb4}} 
\subfigure[]{ \resizebox{4.5cm}{!}{\includegraphics{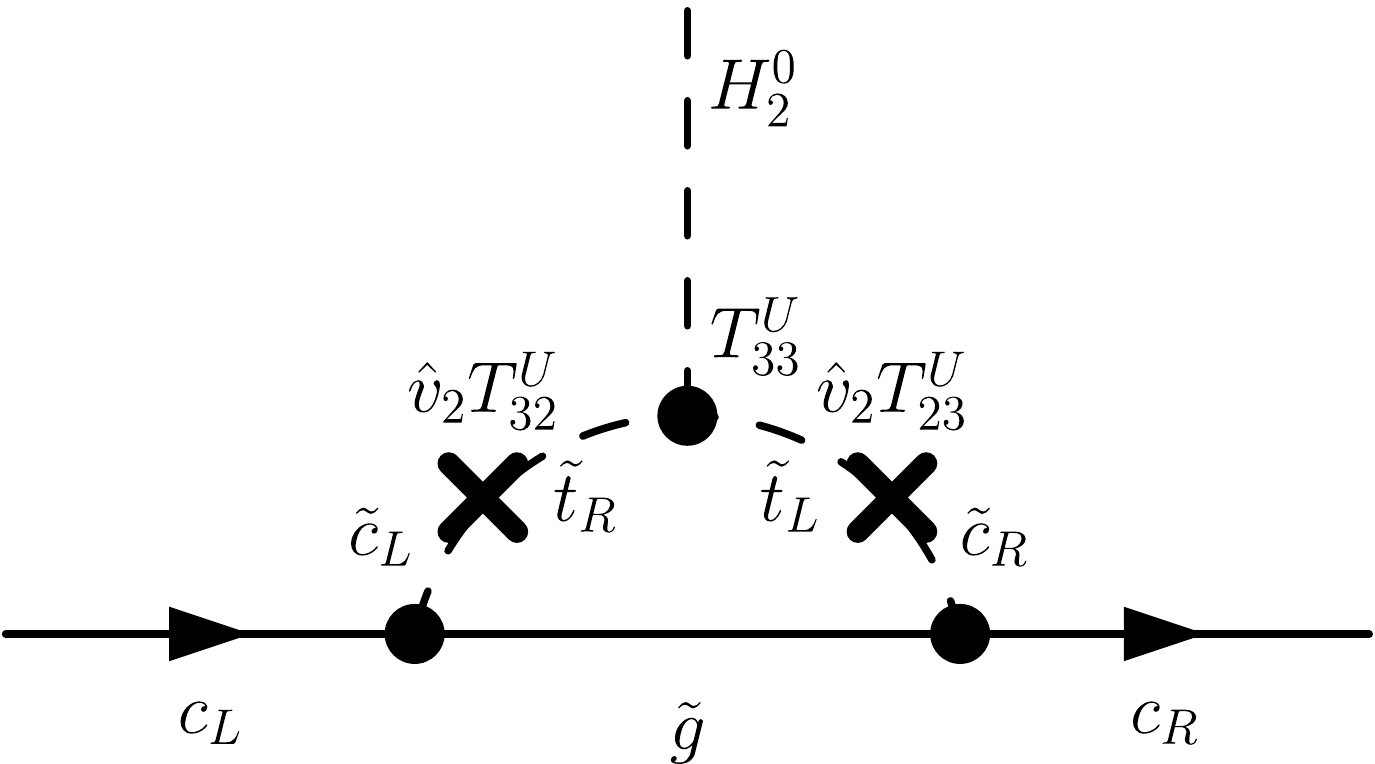}} \label{diag_h0ccb5}} 
\subfigure[]{ \resizebox{4.5cm}{!}{\includegraphics{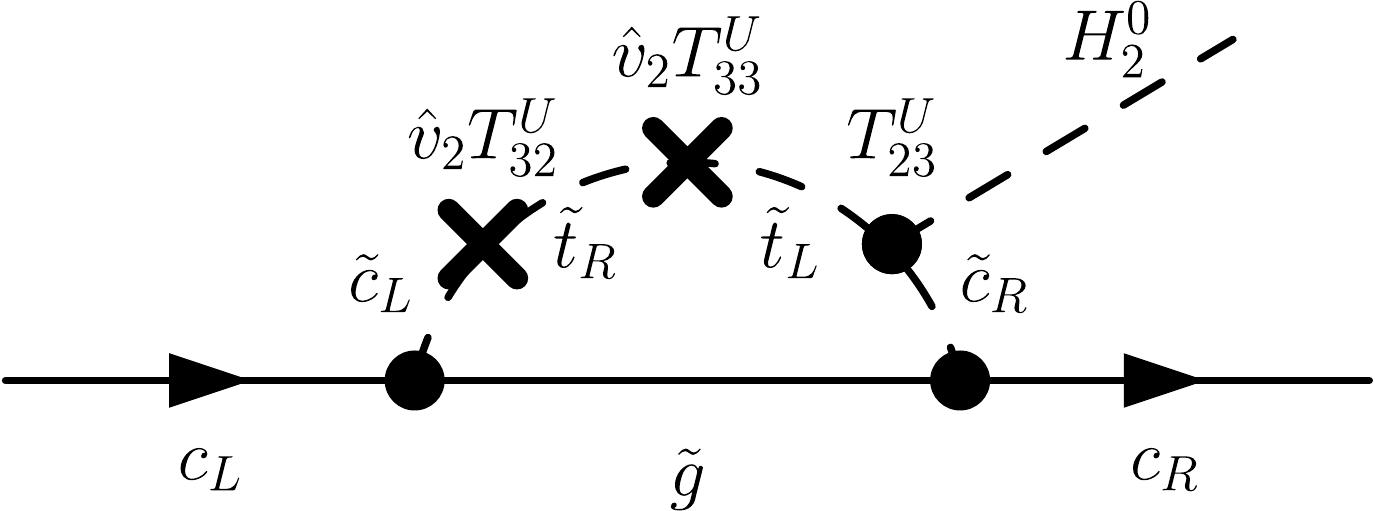}} \label{diag_h0ccb6}} 
\caption{Quark-flavour violating mass insertions to the vertex graph $H^0_2 \to c \bar c$ with $\sg-\su$ loop,
related to $T^v_1$ and $T^v_2$ in eq.~(\ref{MI_Tvs}).}
\label{diag_h0ccb}
\end{figure*}

We now turn to the vertex amplitude of the decay $h^0 \to c\bar c$ with ${\tilde g}$, ${\tilde u}^*_i$ and ${\tilde u}_j$ in the loop,
defined by ${\cal L} = - h^0 \bar c\, (c^v_L P_L + c^v_R P_R)\, c$.  
Neglecting the charm mass and $m_{h^0}$ compared to the gluino and $\su_i$ masses, for the coefficients $c^v_L$ and $c^v_R$ we have 
\begin{eqnarray}
c^v_L & = &  - {2 \alpha_s \over 3 \pi} \sum_{i,j=1}^6 m_{\tilde g}  c_{h^0 \tilde u_i^* \tilde u_j}  U^{\tilde u}_{j 2}  U^{\tilde u *}_{i 5} \,
  c_0(m^2_{\tilde g},  m^2_{\tilde u_i}, m^2_{\tilde u_j})\, ,\\
c^v_R & = & - {2 \alpha_s  \over 3 \pi} \sum_{i,j=1}^6 m_{\tilde g}  c_{h^0 \tilde u_i^* \tilde u_j} U^{\tilde u}_{j 5} U^{\tilde u *}_{i 2} \,
c_0(m^2_{\tilde g},  m^2_{\tilde u_i}, m^2_{\tilde u_j})\, .
\end{eqnarray}
We use $c_0(m^2_{\tilde g},  m^2_{\tilde u_i}, m^2_{\tilde u_j}) = C_0(0, 0, 0, m^2_{\tilde g},  m^2_{\tilde u_i}, m^2_{\tilde u_j})$
with $C_0$ being the scalar Passarino-Veltman integral with three propagators, 
and the coupling $c_{h^0 \tilde u_i^* \tilde u_j}$ is given by eq. (65) of~\cite{Bartl:2014bka}, 
\begin{equation}
c_{h^0 \tilde u_i^* \tilde u_j} = -{\cos\alpha \over \sqrt2} \sum_{l, k = 1,3}\, 
\left(U^{\tilde u *}_{j l} U^{\tilde u}_{i k+3} T^U_{k l} + U^{\tilde u *}_{j l+3} U^{\tilde u}_{i k} T^{U *}_{l k}\right) + \ldots\, .
\end{equation}
Assuming that $T_{U23}, T_{U32}, T_{U33}$ are non-zero and real, we can approximate $c_{h^0 \tilde u_i^* \tilde u_j}$ by
\begin{eqnarray}
c_{h^0 \tilde u_i^* \tilde u_j} & = & - {\cos\alpha \over \sqrt2} \bigg( 
                                       T^U_{23} \left(U^{\tilde u}_{i3} U^{\tilde u *}_{j5} + U^{\tilde u}_{i5} U^{\tilde u *}_{j3}\right)\nonumber\\
&&  \hspace*{1.2cm}  +\,  T^U_{32} \left(U^{\tilde u}_{i2} U^{\tilde u *}_{j6} + U^{\tilde u}_{i6} U^{\tilde u *}_{j2}\right) \nonumber\\
&&   \hspace*{1.2cm} +\, T^U_{33} \left(U^{\tilde u}_{i3} U^{\tilde u *}_{j6} + U^{\tilde u}_{i6} U^{\tilde u *}_{j3}\right) \bigg)\, .
\end{eqnarray}
The mass insertion expansions for the coefficients $c^v_L$ and $c^v_R$, are equal (for real input parameters), $c^v_L = c^v_R = c^v$,
\begin{equation}
c^v=  - \frac{2 \a_s}{3 \pi } \msg  {\cos\alpha \over \sqrt2} (T^v_1 + T^v_2 + \ldots)\, ,
\label{coeff_verth02ccb_MI}
\end{equation}
where
\begin{eqnarray}
T^v_1 & = & T^U_{32} M^U_{23}  \, c_0\! \left(1,  \msg^2, M^{LL}_{22}, \{M^{RR}_{22}, M^{RR}_{33}\} \right) \nn
                    &+&T^U_{23} M^Q_{23} \, c_0\! \left(1,  \msg^2, M^{RR}_{22}, \{M^{LL}_{22}, M^{LL}_{33}\} \right)\,, \nn
T^v_2 & = & T^U_{33} \left(M^Q_{23}  M^U_{23} + 3\, {\hat v_2}^2 \, T^U_{23}  T^U_{32} \right)
\, c_0\! \left(2,  \msg^2, \{ M^{LL}_{22}, M^{LL}_{33} \}, \{M^{RR}_{22}, M^{RR}_{33}\} \right) \,.
\end{eqnarray}
In terms of $b_0$-functions we have
\begin{eqnarray}
T^v_1 & = & T^U_{32} M^U_{23}  \, b_0\! \left(2,  \msg^2, \{ M^{LL}_{22}, M^{RR}_{33}, M^{RR}_{22}\} \right)  \nn
                   &+&T^U_{23} M^Q_{23}  \, b_0\! \left(2,  \msg^2, \{M^{LL}_{22}, M^{LL}_{33}, M^{RR}_{22}\} \right)\,, \nn
T^v_2 & = & T^U_{33} \left(M^Q_{23}  M^U_{23} + 3\, {\hat v_2}^2 \, T^U_{23}  T^U_{32} \right)
\, b_0\! \left(3,  \msg^2, \{ M^{LL}_{22}, M^{LL}_{33}, M^{RR}_{33}, M^{RR}_{22}\} \right) \,. 
\label{MI_Tvs}
\end{eqnarray}
The graphs corresponding to the terms $T^v_1$ and $T^v_2$ are given in Figs.~\ref{diag_h0ccb1} and~\ref{diag_h0ccb2} or~\ref{diag_h0ccb3} to~\ref{diag_h0ccb6}, respectively.
Comparing the results for the charm self-energy, eqs.~(\ref{dsigLRsg_c1}),(\ref{MI_Ts}), and the vertex contribution to  $h^0 \to c \bar c$,
eqs.~(\ref{coeff_verth02ccb_MI}),(\ref{MI_Tvs}), we see that $T_2 = T^v_1 \hat v_2$. The same holds for the term proportional to $T^U_{33} M^Q_{23}  M^U_{23} $
in $T_3$ and $T^v_2$. Concerning the term proportional to $T^U_{33} T^U_{23}  T^U_{32}$ we have a factor 3 in the term  $T^v_2$ compared to that in $T_3$.
This can also be seen by comparing Fig.~\ref{diag_ccb4} with Figs.~\ref{diag_h0ccb4} to~\ref{diag_h0ccb6}. Thus we can deduce the result $T^v_3$ from the term $T_4$ in eq.~(\ref{MI_Ts}) by adding a prefactor of 3 for all the terms with three $T^U$ elements. 

In a recent paper by A. Brignole~\cite{Brignole:2015kva} the width $\Gamma(h^0 \to b \bar{b})$ was also considered in a quark flavour changing scenario. There
only the graphs of Figs.~\ref{diag_h0ccb4} to~\ref{diag_h0ccb6} were taken, which are, however, much suppressed compared to Figs.~\ref{diag_h0ccb1} to~\ref{diag_h0ccb3}.

The leading term in the SUSY contribution to the $\drbar$~$m_c$ is UV-finite and therefore scale independent. 
As $M^Q_{23}$ is strongly constrained by B-physics observables, this term is nearly proportional to the  
product of the two insertions $T^U_{32}$ and $M^U_{23}$, see Fig.~\ref{diag_ccb}. 
The resummed SM running charm mass $m_c|_{SM}$ is
$\sim$ 0.6~GeV. The SUSY $\drbar$ running charm mass can be written then as $m_c \sim 0.6\, {\rm GeV} + \Delta m_c^{\tilde g}$ with 
\begin{equation}
\Delta m_c^{\tilde g} \simeq - {2 \alpha_s \over 3 \pi} m_{\tilde g} \hat v_2 T^U_{32} M^U_{23}  \, b_0\! \left(2,  \msg^2, \{ M^{LL}_{22}, M^{RR}_{33}, M^{RR}_{22}\} \right)\, .
\end{equation}
When all arguments of $b_0(2, \ldots)$ become equal $\sim M_S$, we get 
\begin{equation}
b_0\! \left(2,  M^2_S, \{ M^2_S, M^2_S, M^2_S\} \right) = {1 \over 2 M^4_S}\, ,
\end{equation}
$\hat v_2 \sim 170$~GeV and for $\alpha_s$ we take 0.1. We get
\begin{equation}
\Delta m_c^{\tilde g} \sim - 1.8\, {\rm GeV} \, m_{\tilde g} {T^U_{32} M^U_{23}  \over M^4_S}\, .
\end{equation} 
Let us take $m_{\tilde g} = \sqrt{M^U_{23}} =  M_S$  and $T^U_{32} > M_S/3$. Then the $\drbar$ $m_c \le 0$. 
The product $T^U_{32} M^U_{23}$ can be positive or negative and hence the one-loop 
width is not positive definite. 
In this case perturbation theory is no more valid.

In order to find bounds for $T^U_{32}$ and $M^U_{23}$
we also have studied the decay $t \to c h^0$, having 
written a numerical program for its decay width.
However, the product 
$T^U_{32} M^U_{23}$ cannot be directly constrained by this process. In principle, 
one could get individual bounds on $T^U_{32}$ and $M^U_{23}$ but the effects of these parameters
on the width turn out to be numerically too small~\cite{Dedes:2014asa}.

Neglecting the wave-function contributions, which are proportional to the tree-level coupling $s^1_c$ we get the approximate result for 
the decay $h^0 \to c \bar c$,
\begin{equation}
\Gamma^{\rm appr}(h^0 \to c \bar c) = \Gamma^{g, {\rm impr}} - 2\, \Sigma^{LR, \tilde g}_c\, \Gamma^{\rm tree}(m_c)\,,
\end{equation}
where $\Sigma^{LR, \tilde g}_c$ is given in eq.~(\ref{dsigLRsg_c}) or in the MI approximation in eq.~(\ref{dsigLRsg_c1}) with $\rho = 1$.
$\Gamma^{g, {\rm impr}}$ can be taken from eq.~(55) and $\Gamma^{\rm tree}$ from eq.~(9) in \cite{Bartl:2014bka}. 
\begin{figure*}[h!]
\centering
\subfigure[]{ \resizebox{4.5cm}{!}{\includegraphics{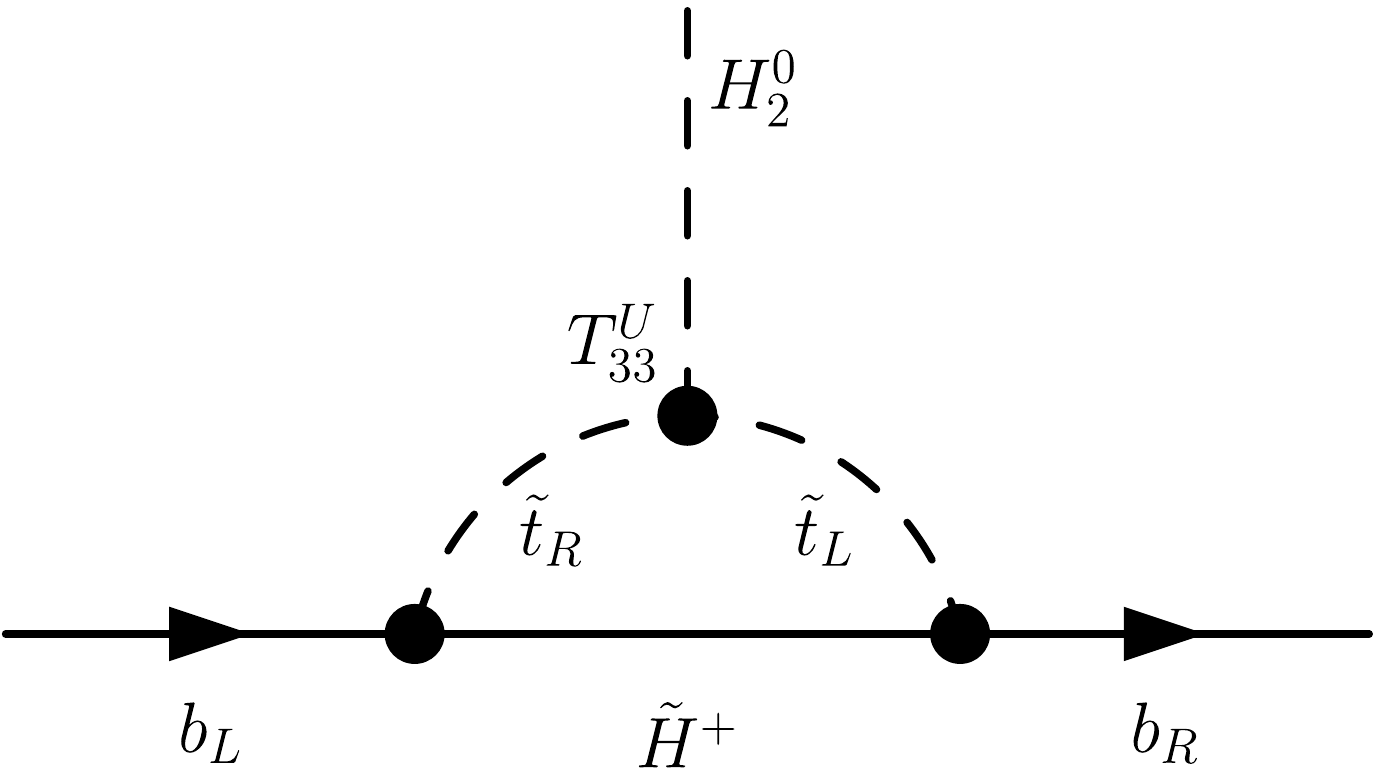}} \label{diag_h0bbb1}}
\subfigure[]{ \resizebox{4.5cm}{!}{\includegraphics{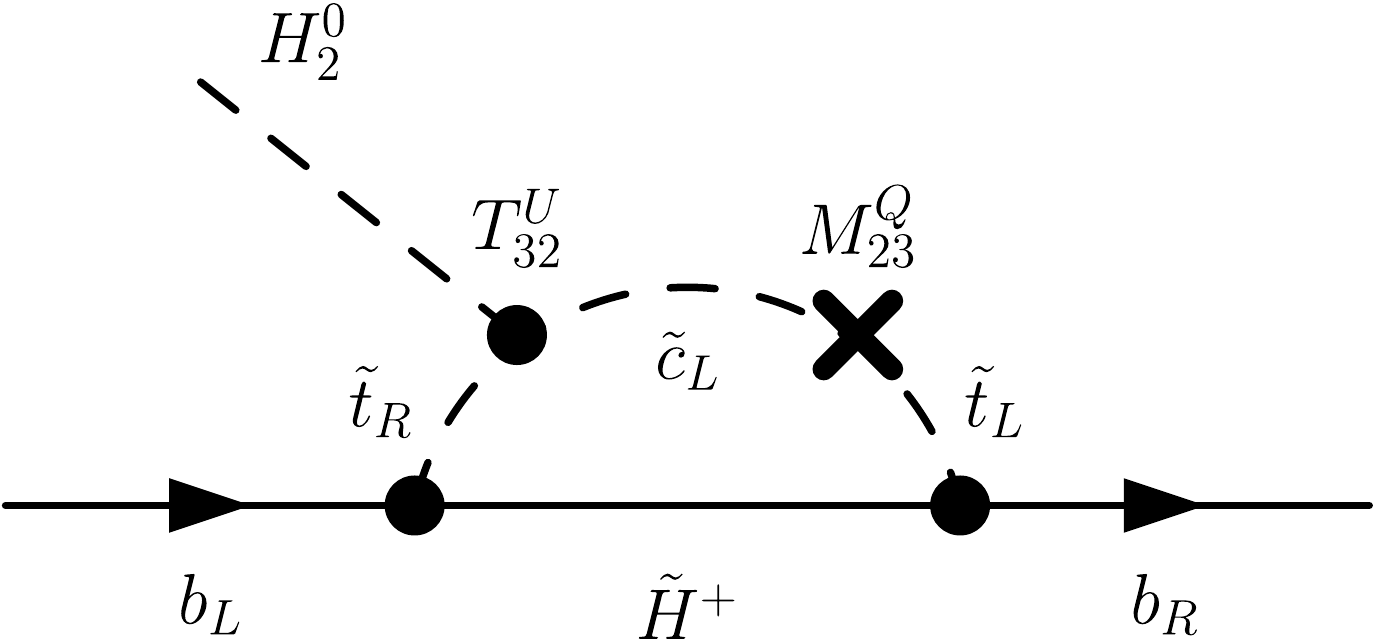}} \label{diag_h0bbb2}} 
\subfigure[]{ \resizebox{4.5cm}{!}{\includegraphics{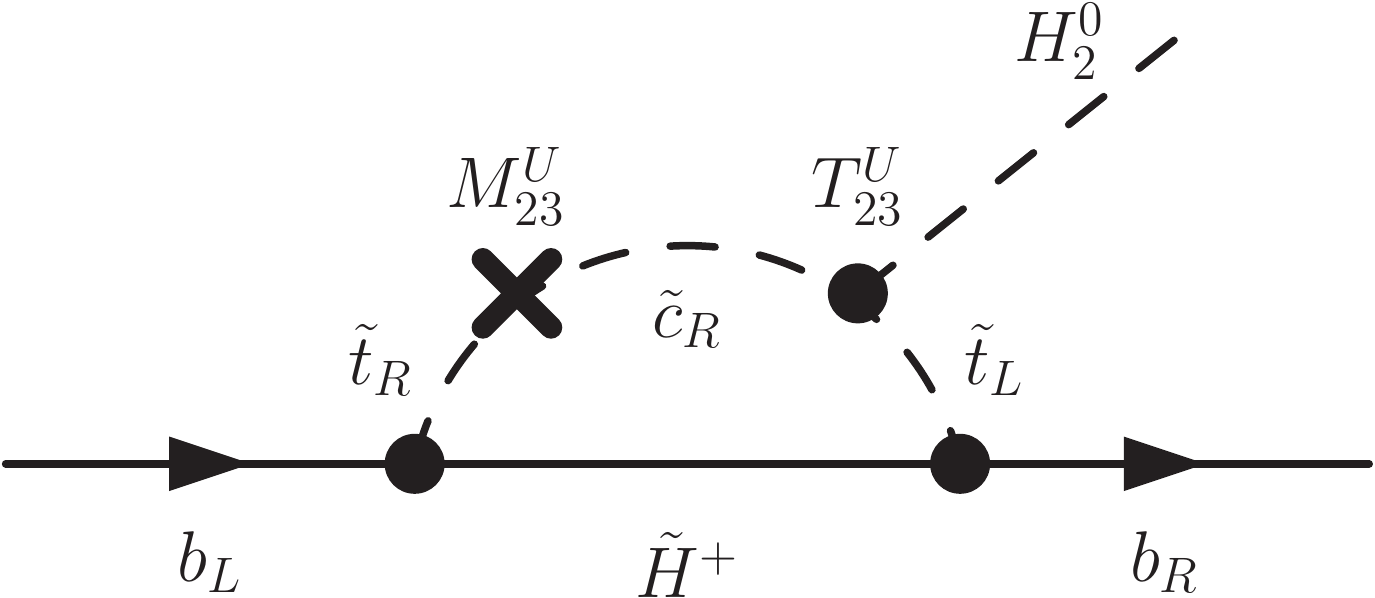}} \label{diag_h0bbb3}} \\
\subfigure[]{ \resizebox{4.5cm}{!}{\includegraphics{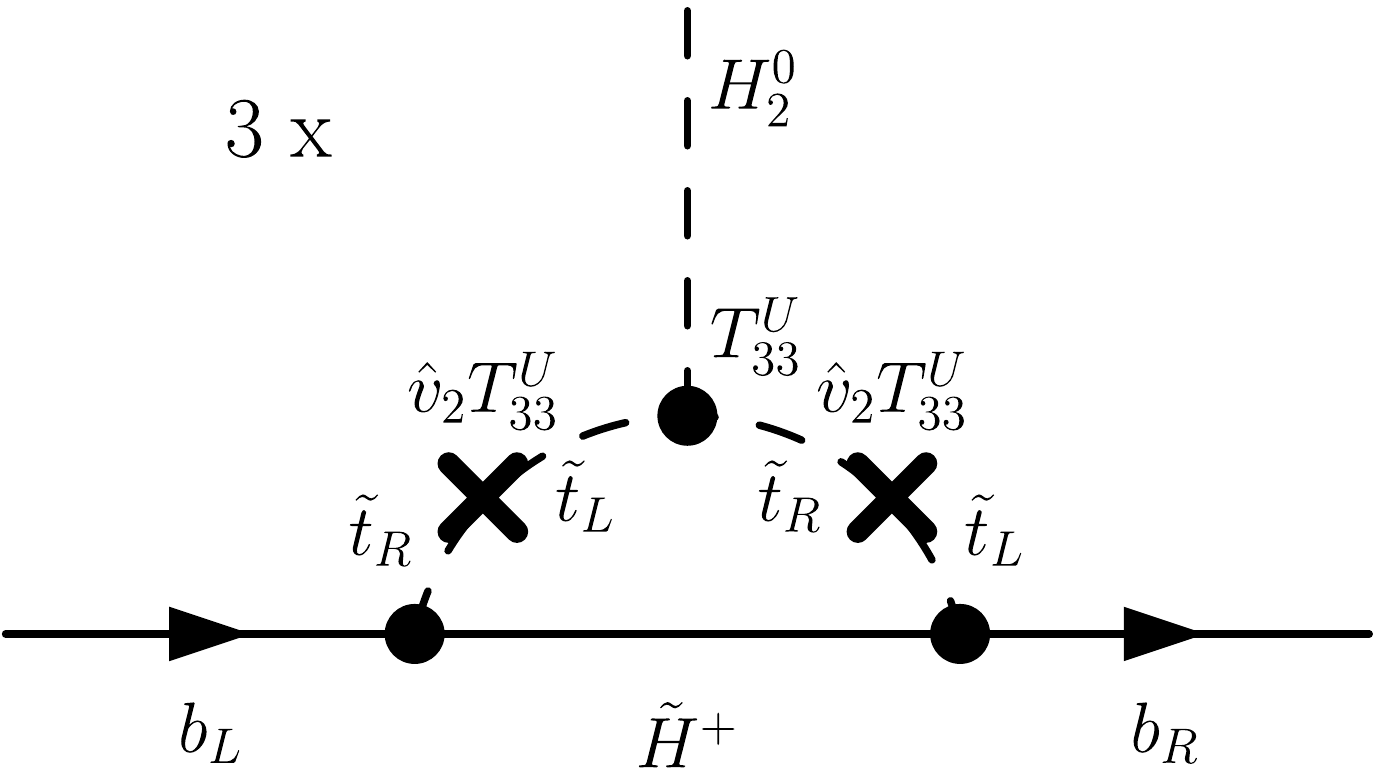}} \label{diag_h0bbb4}} 
\subfigure[]{ \resizebox{4.5cm}{!}{\includegraphics{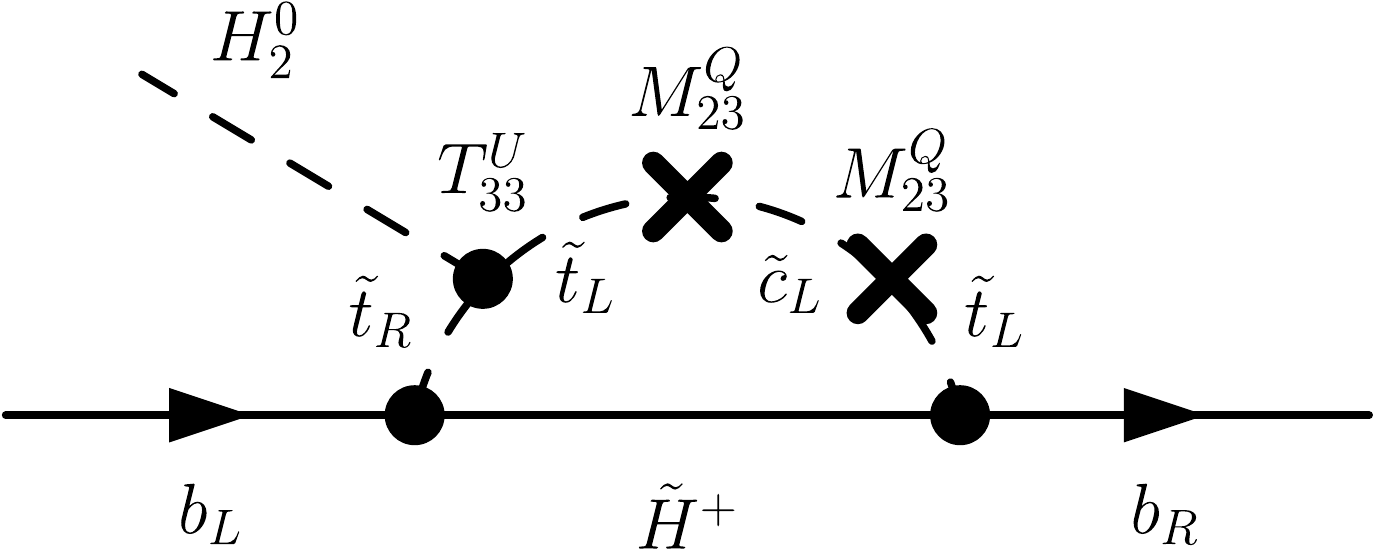}} \label{diag_h0bbb5}} 
\subfigure[]{ \resizebox{4.5cm}{!}{\includegraphics{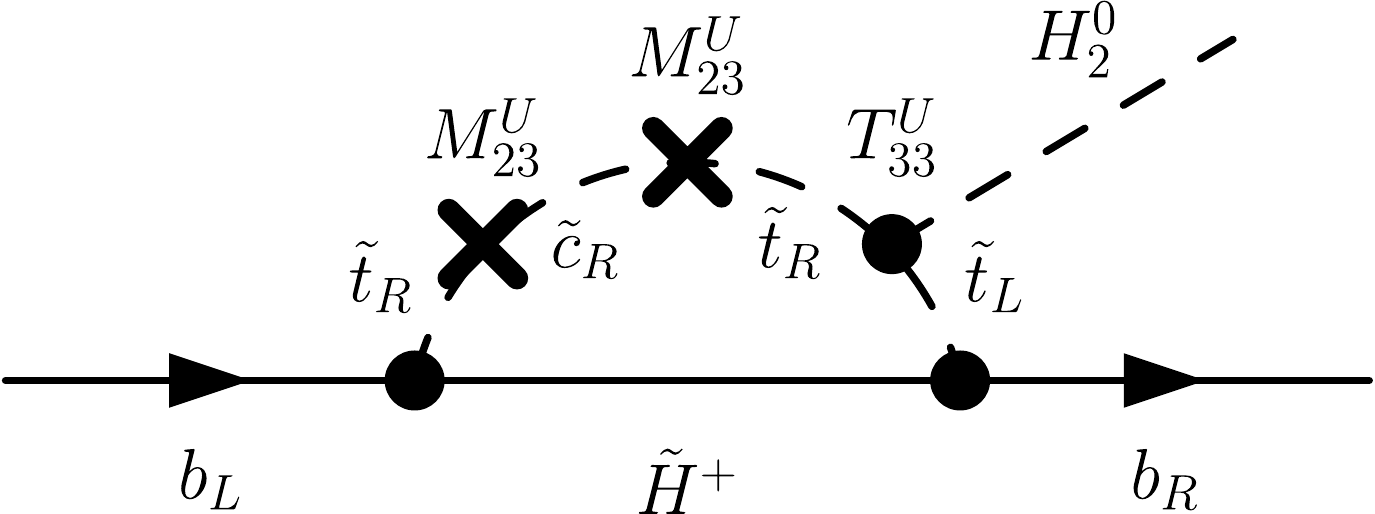}} \label{diag_h0bbb6}} \\
\subfigure[]{ \resizebox{4.5cm}{!}{\includegraphics{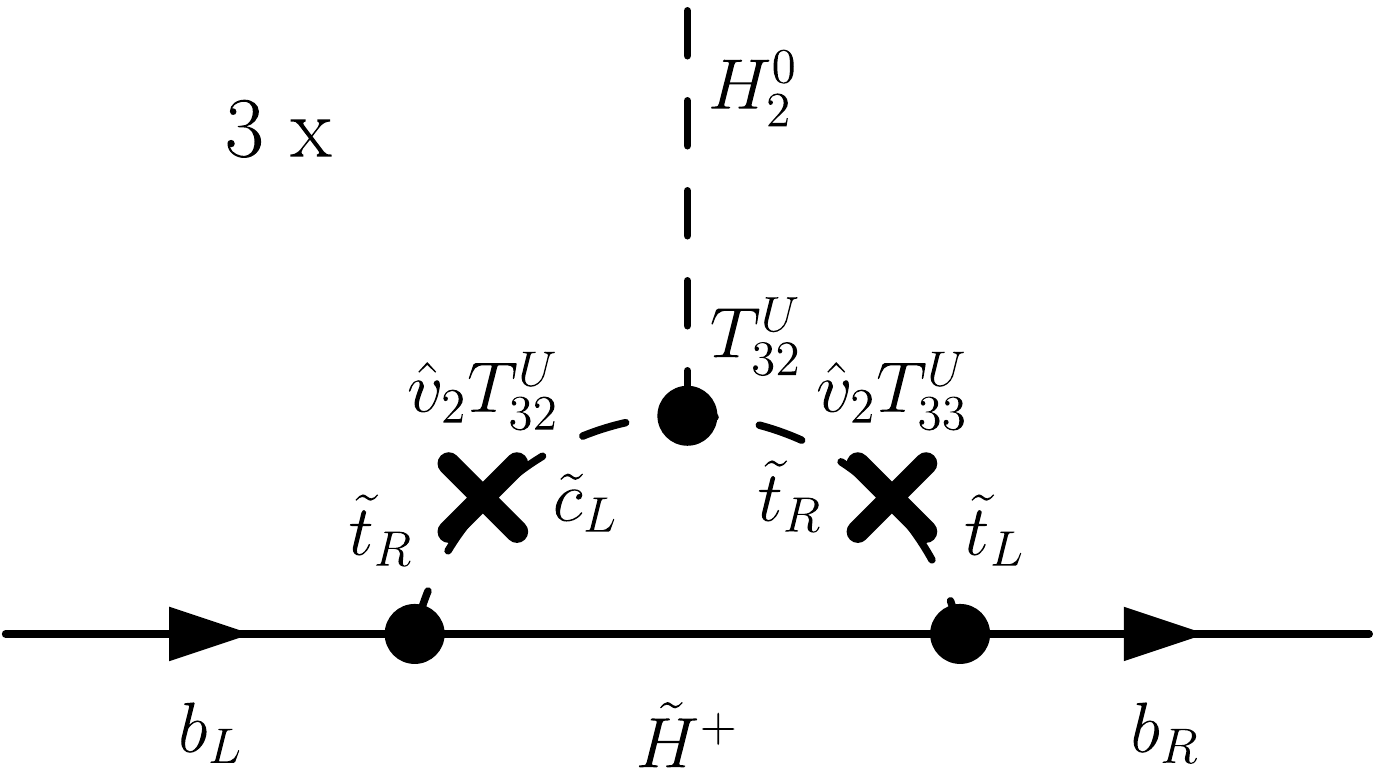}} \label{diag_h0bbb7}} 
\subfigure[]{ \resizebox{4.5cm}{!}{\includegraphics{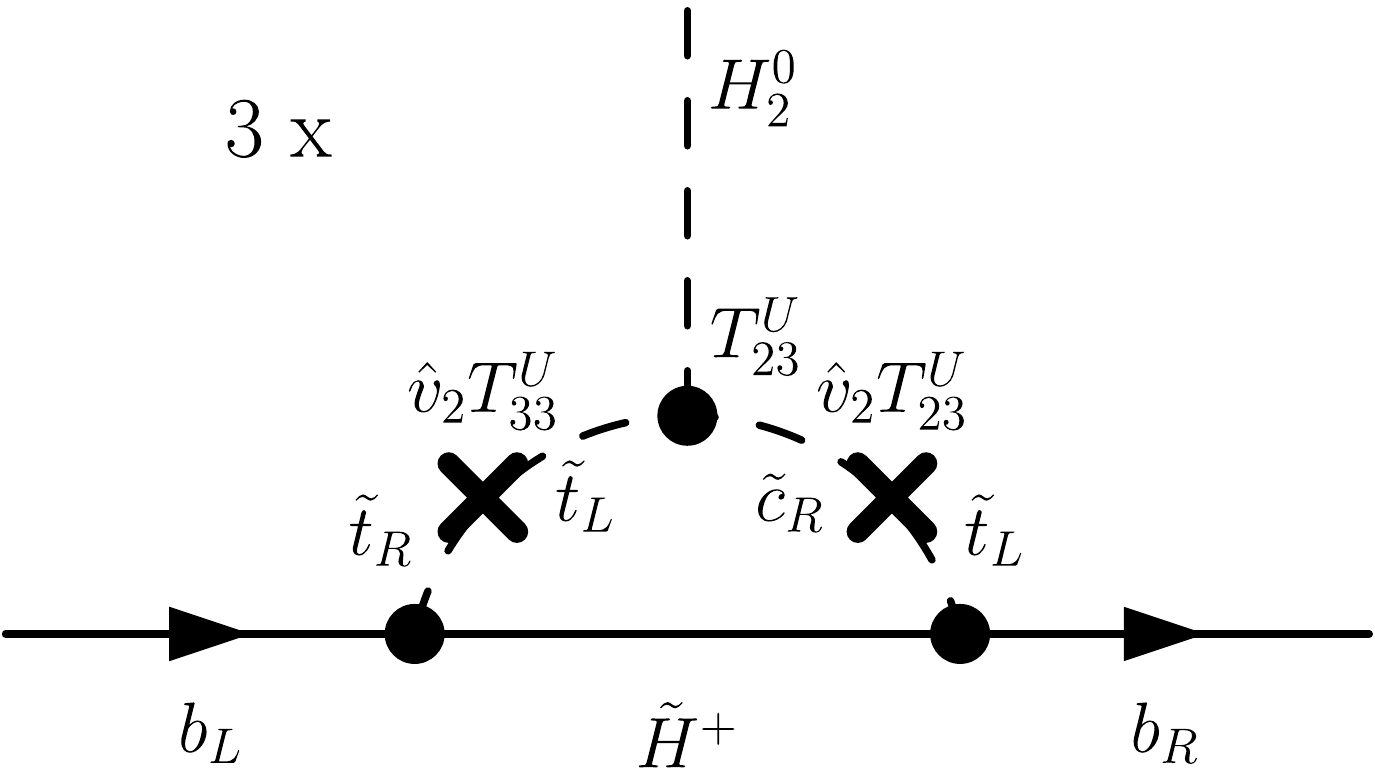}} \label{diag_h0bbb8}} 
\caption{Quark-flavour violating mass insertions to the vertex graph  $H^0_2 \to b \bar b$ with charged higgsino loop,
proportional to $h_b h_t$, 
see eqs.~(\ref{mbchaMI2}) and (\ref{mbchaMI1Ts}).}
\label{diag_h0bbb}
\end{figure*}

\subsection{Gluino and chargino contributions to \boldmath $h^0 \to b \bar b$}
\label{SecMI:gluinocha2bbb}

We decompose the bottom self energy $\Sigma_b$ defined by the Lagrangian
${\cal L} = - \bar b\, \Sigma_b\, b$ as follows
\begin{equation}
\Sigma_b( p ) = \slashed{p} \left( \Sigma^{LL}_b(p^2) P_L + \Sigma^{RR}_b(p^2) P_R \right) +
m_b \left( \Sigma^{RL}_b(p^2) P_L + \Sigma^{LR}_b(p^2) P_R \right)  \, ,
\label{sigma_decomp_b}
\end{equation}
with $\Sigma^{LR}_b = \Sigma^{RL *}_b$. 
We have $\Sigma^{LR}_b = \Sigma^{RL}_b$, as we assume real input parameters, and 
\begin{equation}
m_{b} \Sigma^{LR, \tilde g}_b =   - \frac{2 \a_s}{3 \pi } \msg\,\sum_{i=1}^6 U^{\sd *}_{i3}U^{\sd}_{i6}
B_0(m_b^2, m^2_{ \tilde g}, m^2_{\tilde d_i})\, .
\label {dsigLRsg_b}
\end{equation}
Allowing the squared $\tilde d$-mass matrix in the form
\begin{equation}
\hspace*{-1cm}    {\cal M}^2_{\tilde{d}} = \left( \begin{array}{cc}
        {\cal M}^2_{\tilde{d},LL} & {\cal M}^2_{\tilde{d},LR} \\[2mm]
        {\cal M}^2_{\tilde{d},RL} & {\cal M}^2_{\tilde{d},RR} \end{array} \right) \equiv M_{ij} =          
         \left( \begin{array}{cccccc}
         M^{LL}_{11} & 0 & 0 & 0 & 0 & 0 \\
         0  & M^{LL}_{22} & M^Q_{23} & 0 & 0 & \hat v_1 T^D_{32} \\
         0 & M^Q_{23} & M^{LL}_{33} & 0 &  \hat v_1 T^D_{23} & M^{RL}_{33} \\
         0 & 0 & 0 &  M^{RR}_{11} & 0 & 0\\
         0 & 0 &   \hat v_1 T^D_{23} & 0 & M^{RR}_{22} &  M^D_{23} \\
         0 & \hat v_1 T^D_{32} & M^{RL}_{33}  & 0 &  M^D_{23} & M^{RR}_{33}
         \end{array}\right)  \, ,     
 \label{EqMassMatrix_sd}
\end{equation}
with $M^{RL}_{33} \sim  - \mu m_b \tan\beta$, $\hat v_1 = v \cos\beta/\sqrt{2} \sim $ 170~GeV$/\tan\beta$, and the QFV elements of the  $3 \times 3$ matrices $M^2_Q$ and $M^2_D$ are
written as $M^Q_{ij}$ and $M^D_{ij}$, respectively. Using eq.~(\ref{mass2MI_exp4}) we get 
\begin{equation}
m_b \Sigma^{LR, \tilde g}_b =   - \frac{2 \a_s}{3 \pi } \msg (T^{FC}_1 + T^{FV}_2 + T^{FC}_3  + T^{FV}_3 + \ldots)
\label {dsigLRsg_b1}
\end{equation}
where the quark flavour conserving (FC) and quark flavour violating (FV) contributions read
\begin{eqnarray}
T^{FC}_1 & = & M^{RL}_{33}  \, b_0\! \left(1,  \msg^2, \{ M^{RR}_{33}, M^{LL}_{33}\} \right)\nn
T^{FV}_2 & = & \hat v_1 T^D_{32} M^Q_{23}  \, b_0\! \left(2,  \msg^2, \{ M^{RR}_{33}, M^{LL}_{22}, M^{LL}_{33}\} \right) \nn
              &+&   \hat v_1 T^D_{23} M^D_{23}  \, b_0\! \left(2,  \msg^2, \{M^{RR}_{33}, M^{RR}_{22}, M^{LL}_{33}\} \right) \nn
T^{FC}_3 & = & 3^\rho\, (M^{RL}_{33})^3   \, b_0\! \left(3,  \msg^2, \{ M^{RR}_{33}, M^{LL}_{33}, M^{RR}_{33}, M^{LL}_{33}\} \right)\nn
T^{FV}_3 & = & (M^Q_{23})^2  M^{RL}_{33}   \, b_0\! \left(3,  \msg^2, \{ M^{RR}_{33}, M^{LL}_{33}, M^{LL}_{22}, M^{LL}_{33}\} \right)\nn
        & + &  (M^D_{23})^2  M^{RL}_{33}   \, b_0\! \left(3,  \msg^2, \{ M^{RR}_{33}, M^{RR}_{33}, M^{RR}_{22}, M^{LL}_{33}\} \right)\nn
       & + & 3^\rho\, (\hat v_1)^2 (T^D_{32})^2  M^{RL}_{33} \, b_0\! \left(3,  \msg^2, \{ M^{RR}_{33}, M^{LL}_{22}, M^{RR}_{33}, M^{LL}_{33}\} \right)\nn      
       & + & 3^\rho\, (\hat v_1)^2  (T^D_{23})^2  M^{RL}_{33} \, b_0\! \left(3,  \msg^2, \{ M^{RR}_{33}, M^{LL}_{33}, M^{RR}_{22}, M^{LL}_{33}\} \right)\, ,
\end{eqnarray}     
with $\rho = 0$.
As in the charm sector, the vertex contribution can be directly deduced from the self~energy,  $T^{v\, x}_i = {T^{x}_i \over \hat v_1}, x = {FC, FV}$, 
with an additional factor 3 for some terms in $T^x_3$,  $(M^{RL}_{33})^3  \to 3 (M^{RL}_{33})^3$, $(T^D_{23})^2 M^{RL}_{33} \to 3 (T^D_{23})^2 M^{RL}_{33}$,
$(T^D_{32})^2 M^{RL}_{33} \to 3 (T^D_{32})^2 M^{RL}_{33}$, and accordingly $\rho = 1$.
The  interactions related to the mass insertions are given by
\bea
\hspace*{-1cm}{\cal L} &=& -T^D_{33}\, {\tilde b}_R^{*}  {\tilde b}_L H^0_1 - T^D_{32}\,  {\tilde b}_R^{*}  {\tilde s}_L H^0_1
- T^D_{23}\, {\tilde s}_R^{*} {\tilde b}_L H^0_1 - M^{RL}_{33} \, {\tilde b}_L^{*} {\tilde b}_R - M^Q_{23}\, {\tilde b}_L^{*} {\tilde s}_L \nn
&&  - M^D_{23}\, {\tilde b}_R^{*} {\tilde s}_R + {\rm h.c.}\,,
\label{}
\eea
with  $H^0_1 = {1 \over \sqrt2}(v \cos\beta - h^0 \sin\alpha  + \ldots)$.\\

We will apply the mass insertion technique for the self-energy amplitude
of the bottom-quark and for the vertex amplitude with a chargino in the loop.
The relevant term for the self energy calculation
is proportional to $c^*_L c_R$, with $c_L = h_b U^*_{m 2} U^{\tilde u *}_{i 3}$ and
$c_R = - g V_{m 1} U^{\tilde u *}_{i 3} + h_t V_{m 2} U^{\tilde u *}_{i 6}$. Using eq. (\ref{sigma_decomp_b}) we get
\begin{eqnarray}
\hspace*{-0.7cm}
m_b \Sigma^{LR, \tilde \chi^+}_b & =  & {1 \over 16 \pi^2} \sum_{m = 1}^2 \sum_{i = 1}^6 
m_{\tilde \chi^+_m} \left( - g h_b U_{m 2} V_{m 1} |U^{\tilde u}_{i 3}|^2 + 
h_b h_t U_{m 2} V_{m 2}  U^{\tilde u}_{i 3} U^{\tilde u *}_{i 6} \right) \times \nonumber \\
&& \hspace*{7cm}B_0(m_b^2, m^2_{ \tilde \chi^+_m}, m^2_{\tilde u_i})\, .
\label{sigLRbottom}
\end{eqnarray}
Neglecting the term
proportional to the SU(2) coupling~$g$ and the bottom mass in the loop integrals, we get
\begin{equation}
m_b \Sigma^{LR, \tilde \chi^+}_b =  {h_b h_t  \over 16 \pi^2} \sum_{m = 1}^2 \sum_{i = 1}^6  m_{\tilde \chi^+_m}  U_{m 2} V_{m 2}\, 
U^{\tilde u}_{i 3} U^{\tilde u *}_{i 6} b_0(m^2_{ \tilde \chi^+_m}, m^2_{\tilde u_i})\, .
\end{equation} 
Concerning the mass insertions in the $\tilde u_i$ line, we have the same structure as in eq.~(\ref{dsigLRsg_b}), but
for the $\su$ sector. We have $M^{RL}_{33}  \to \hat v_2 T^U_{33}$, $T^D \to T^U$, and $M^D_{23} \to M^U_{23}$.
Therefore, we can use the results for the bottom self energy with gluino in the loop. Using eq.~(\ref{EqMassMatrix}) we obtain
\begin{equation}
m_b \Sigma^{LR, \tilde \chi^+}_b =  {h_b h_t  \over 16 \pi^2} \sum_{m = 1}^2 m_{\tilde \chi^+_m}  U_{m 2} V_{m 2}\, 
\left(T^{FC}_1 + T^{FV}_2 + T^{FC}_3  + T^{FV}_3 + \ldots\right)\, ,
\label{mbchaMI1}
\end{equation} 
where
\begin{eqnarray}
T^{FC}_1 & = &  \hat v_2 T^U_{33}  \, b_0\! \left(1,  m_{\tilde \chi^+_m}^2, \{ M^{RR}_{33}, M^{LL}_{33}\} \right)\nn
T^{FV}_2 & = & \hat v_2 T^U_{32} M^Q_{23}  \, b_0\! \left(2,  m_{\tilde \chi^+_m}^2, \{ M^{RR}_{33}, M^{LL}_{22}, M^{LL}_{33}\} \right)\nn 
                & + &   \hat v_2 T^U_{23} M^U_{23}  \, b_0\! \left(2, m_{\tilde \chi^+_m}^2, \{M^{RR}_{33}, M^{RR}_{22}, M^{LL}_{33}\} \right) \nn
T^{FC}_3 & = & 3^\rho\, ( \hat v_2 T^U_{33})^3   \, b_0\! \left(3, m_{\tilde \chi^+_m}^2, \{ M^{RR}_{33}, M^{LL}_{33}, M^{RR}_{33}, M^{LL}_{33}\} \right)\nn
T^{FV}_3 & = &   \hat v_2 T^U_{33}  (M^Q_{23})^2   \, b_0\! \left(3,  m_{\tilde \chi^+_m}^2, \{ M^{RR}_{33}, M^{LL}_{33}, M^{LL}_{22}, M^{LL}_{33}\} \right)\nn
        & + &  \hat v_2 T^U_{33} (M^U_{23})^2  \, b_0\! \left(3,  m_{\tilde \chi^+_m}^2, \{ M^{RR}_{33}, M^{RR}_{33}, M^{RR}_{22}, M^{LL}_{33}\} \right)\nn
       & + & 3^\rho\, (\hat v_2)^3 (T^U_{32})^2\,  T^U_{33}   \, b_0\! \left(3, m_{\tilde \chi^+_m}^2, \{ M^{RR}_{33}, M^{LL}_{22}, M^{RR}_{33}, M^{LL}_{33}\} \right)\nn      
       & + & 3^\rho\, (\hat v_2)^3  (T^U_{23})^2\,  T^U_{33}   \, b_0\! \left(3,  m_{\tilde \chi^+_m}^2, \{ M^{RR}_{33}, M^{LL}_{33}, M^{RR}_{22}, M^{LL}_{33}\} \right)\, ,
\label{mbchaMI1Ts}
\end{eqnarray} 
with $\rho = 0$. The graphs corresponding to $T_{1}$, $T_{2}$ and $T_{3}$ are shown in Figs.~\ref{diag_h0bbb1}, (\ref{diag_h0bbb2},~\ref{diag_h0bbb3}) and (\ref{diag_h0bbb4}-\ref{diag_h0bbb8}), respectively. 
Furthermore, we also apply the mass insertion technique to the chargino part in eq.~(\ref{mbchaMI1}). The eigenvalue equation 
is $U^* X V^{-1} = M_D = {\rm diag}(m_{\tilde \chi^+_1}, m_{\tilde \chi^+_2})$. We assume the chargino mass matrix to be real,
\begin{eqnarray}
X & = & \left( \begin{array}{cc} M_2 & \sqrt2 m_W \sin\beta \\  \sqrt2 m_W \cos\beta & \mu \end{array} \right)\,, \\
\XX \equiv X^\dagger X & = & 
\left( \begin{array}{cc}  M_2^2 + 2 m^2_W \cos^2\beta & \sqrt2 m_W ( \mu \cos\beta  + M_2 \sin\beta) \\  
                \sqrt2 m_W ( \mu \cos\beta  + M_2 \sin\beta) & \mu^2 + 2 m^2_W \sin^2\beta\end{array} \right)\,. \nonumber
\end{eqnarray}
The formula with linear mass insertion reads
\begin{equation}
 \sum_{m = 1}^2 m_{\tilde \chi^+_m}  U_{m 2} V_{m 2}\,  f(m^2_{\tilde \chi^+_m})  =
 X_{22} f(\XX_{22})  +  X_{21} \XX_{21} {f(\XX_{22}) - f(\XX_{11}) \over \XX_{22} - \XX_{11}}\, .
\end{equation}
Assuming $m_W \ll M_2, \mu$, the linear term vanishes, and $X_{22} f(\XX_{22})  \sim \mu f(\mu^2)$. We get the final approximate
result
\begin{equation}
m_b \Sigma^{LR, \tilde \chi^+}_b =  {h_b h_t  \over 16 \pi^2} \mu\,  
\left(T^{FC}_1 + T^{FV}_2 + T^{FC}_3  + T^{FV}_3 + \ldots\right)\, ,
\label{mbchaMI2}
\end{equation} 
with the terms $T^x_{i}$ taken from eq.~(\ref{mbchaMI1Ts}) with $m_{\tilde \chi^+_m}^2 \to \mu^2$.

Neglecting the wave-function renormalization contributions, which are proportional to the tree-level coupling $s^1_b$ we get the approximate result for 
the decay $h^0 \to b \bar b$,
\begin{equation}
\Gamma^{\rm appr}(h^0 \to b \bar b) = \Gamma^{g, {\rm impr}} - 2 \left( \Sigma^{LR, \tilde g}_b +  
\Sigma^{LR, \tilde \chi^+}_b\right)\, \Gamma^{\rm tree}(m_b)\,,
\label{MIapprox}
\end{equation}
where $\Sigma^{LR, \tilde g}_b$ and $\Sigma^{LR, \tilde \chi^+}_b$ are given in eq. (\ref{dsigLRsg_b}) and eq. (\ref{sigLRbottom}) or in the MI-approximation in eq.  (\ref{dsigLRsg_b1}) and eq. (\ref{mbchaMI2}), 
respectively, with $\rho = 1$.
$\Gamma^{g, {\rm impr}}$ is given by eq.~(55) and $\Gamma^{\rm tree}$ by eq.~(9) in \cite{Bartl:2014bka}, with $c \to b$.  
 
In~\cite{Crivellin:2010er} the chirally enhanced corrections to Higgs vertices in the most general MSSM 
were discussed analytically by taking into account gluino-squark loops. We qualitatively
agree with their results on $h^0 \to b \bar b$. A study including two-loop SUSY-QCD corrections
was performed in~\cite{Crivellin:2012zz}.

\section{Numerical results}
\label{sec:num}

In this section we demonstrate the effects of QFV due to $\sca-\st$ mixing in the decays of $h^0$ to $b\bar{b}$ and $c\bar{c}$ in the MSSM.\footnote{In the  $b\bar{b}$ case there are one-loop diagrams with gluino (neutralino) and down-type squark exchange with $\ss_{L,R}-\sb_{L,R}$ mixing. The $\ss_{L}-\sb_{R}$ and $\ss_{R}-\sb_{L}$ mixing is, however, strongly constrained by the vacuum stability conditions~\cite{Bartl:2014bka}, and in addition proportional to $v_1\sim v/\tan \b$, which results in very small $\ss-\sb$ mixing effect. Therefore $\ss-\sb$ mixing will be neglected in our analysis. } 
In order to find an explicit scenario where both decay widths deviate appreciably from the SM values, we have performed two scans over wide parameter regions. 
In the first calculation we have scanned 8750000 parameter points. From them only 17\% have satisfied the existing theoretical and experimental constraints (see Appendix~\ref{sec:constr}). The parameters involved and their variations are given as follows:
\begin{eqnarray}
\{M^2_{U11}, M^2_{U22}, M^2_{U33}\}~[\gev^2] & =&~{\rm in~sets~of}~\bigg\{\{2400^2, 2300^2, 1800^2\},\nonumber \\
&&  \{3000^2, 2800^2, 2000^2\}, \{3200^2, 3000^2, 2200^2\}, \nonumber \\
&& \{2400^2,1100^2,1000^2\}, \{3200^2, 2200^2, 2000^2\}\bigg\}; \nonumber \\
\{M^2_{Q11}, M^2_{Q22}, M^2_{Q33}\}& =& \{M^2_{U11}, M^2_{U22}, M^2_{U33}\};  \nonumber \\
\tan \beta & =&  \{15 \div 30\}~{\rm with~step~size}~2.5; \nonumber \\
\mu ~ [\gev]  & =&  \{1200 \div 2200\} ~{\rm with~step~size}~250; \nonumber \\
\{M_1, M_2, M_3\} ~ [\gev] & =& ~{\rm in~sets~of}~\bigg\{\{300, 600, 1800\}, \{400, 800, 2000\}, \nonumber \\
&&\{500,1000, 2200\}, \{600,1200, 2200\}, \nonumber \\
&&\{700,1400,2400\}\bigg\} \nonumber; \nonumber \\
M^2_{U23}  ~[\gev^2] & =&  \{-2430^2 \div 2430^2\} ~{\rm with~step~size}~\approx1\times10^6;  \nonumber \\
M^2_{Q23}  ~[\gev^2] & =&   \{-1140^2 \div 1140^2 \} ~{\rm with~step~size}~\approx2.9\times10^5; \nonumber \\
T_{U23} ~ [\gev]& =&  \{-3000 \div 3000\} ~{\rm with~step~size}~400; \nonumber \\
T_{U32} ~ [\gev]& =& \{-3000 \div 3000\} ~{\rm with~step~size}~400.
\label{scan1}
\end{eqnarray}
In the second calculation we have varied in more detail the parameters of the mass matrices $M_U$ and $M_Q$, which in the first step have been assumed to change only simultaneously in sets of equal diagonal elements, $(M_U)_{ii} = (M_Q)_{ii}$, for $i=1,2,3$. In this calculation we have scanned 9834496 points and 12\% of them have survived the constraints. The parameters involved and their variations are given by:
\begin{eqnarray}
M^2_{U22}~[\gev^2] & =& \{1000^2 \div 3200^2\}~{\rm with~step~size}~\approx1.3\times 10^6; \nonumber \\%
M^2_{U33} ~[\gev^2] & =&  \{970^2 \div 3100^2\}~{\rm with~step~size}~1.2\times 10^6; \nonumber \\%
M^2_{Q22}~[\gev^2] & =& \{950^2 \div 3150^2\}~{\rm with~step~size}~\approx1.5\times 10^6; \nonumber \\%
M^2_{Q33} ~[\gev^2] & =&  \{1100^2 \div 3100^2\}~{\rm with~step~size}~1.4\times 10^6; \nonumber \\%
M^2_{U23}  ~[\gev^2] & =&  \{-2400^2 \div 2400^2\} ~{\rm with~step~size}~\approx1.6\times10^6;  \nonumber \\
M^2_{Q23}  ~[\gev^2] & =&   \{-1150^2 \div 1150^2\} ~{\rm with~step~size}~\approx4.4\times10^5; \nonumber \\
T_{U23} ~[\gev]& =&  \{-3000 \div 3000\} ~{\rm with~step~size}~\approx1\times 10^3; \nonumber \\
T_{U32} ~[\gev]& =& \{-3000 \div 3000\} ~{\rm with~step~size}~\approx850.
\label{scan2}
\end{eqnarray}
In both scans the following parameters have been fixed:
$M^2_{D11}=1.024\times 10^7~\gev^2, M^2_{D22}=9\times 10^6~\gev^2,M^2_{D33}\approx 7\times 10^6~\gev^2, T_{U33}= 2000~\gev,  
M^2_{D23}=M^2_{D32}=T_{D23}=T_{D32}=T_{D33}=0$.
In the second scan we have also fixed the parameters: $ M_1 = 400~\gev, M_2 = 800~\gev, M_3 = 2000~\gev,
\mu = 1800~\gev, \tan\beta=30, M^2_{U11}=M^2_{Q11}=1.024\times 10^7~\gev^2.$ 
A detailed study of the MSSM QFV parameter space has also been done in~\cite{DeCausmaecker:2015yca}.

\begin{figure*}[h!]
\centering
\subfigure[]{
   { \mbox{\hspace*{-1cm} \resizebox{7.5cm}{!}{\includegraphics{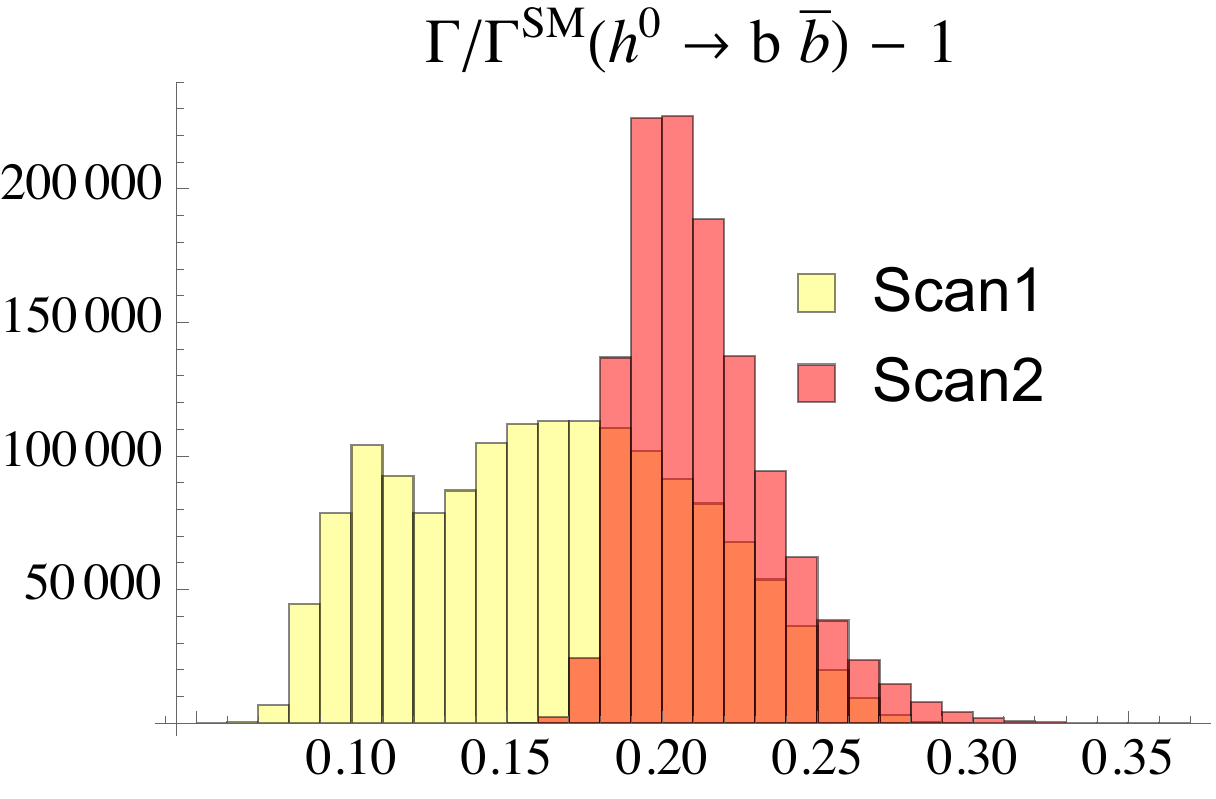}} \hspace*{-0.8cm}}}
   \label{fig1a}}
 \subfigure[]{
   { \mbox{\hspace*{+0.cm} \resizebox{7.5cm}{!}{\includegraphics{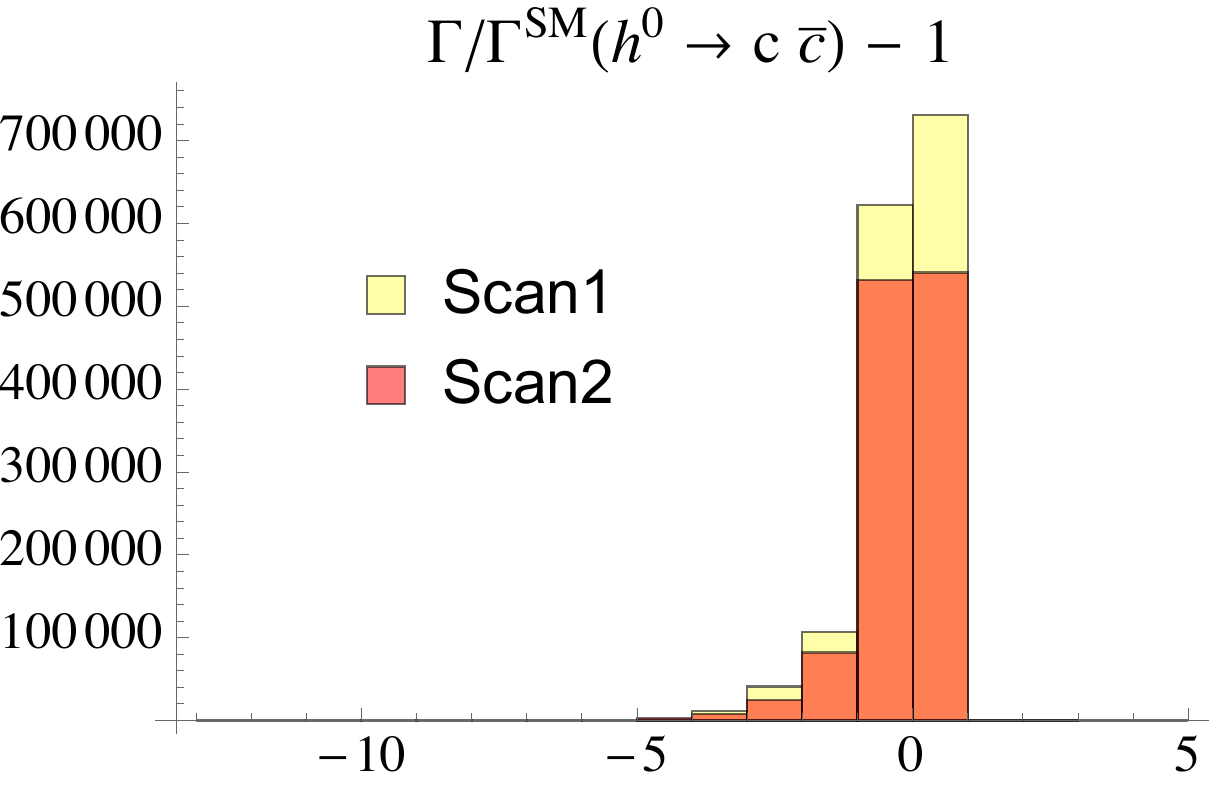}} \hspace*{-1cm}}}
  \label{fig1b}}\\
\caption{Distribution of the results for the deviation from the SM (a)~$\Gamma$/$\Gamma^{\rm SM}(h^0 \to b \bar{b})-1$ and (b)~$\Gamma$/$\Gamma^{\rm SM}(h^0 \to c \bar{c})-1$ from Scan 1 and Scan 2.
}
\label{fig1}
\end{figure*}

The results of the scans are summarised in Fig.~\ref{fig1},
where the distributions of the deviation from the SM width for $h^0 \to b \bar{b}$ and $h^0 \to c \bar{c}$ are shown. We take $\Gamma^{\rm SM}(h^0 \to b \bar{b}) =2.35$ MeV~\cite{Almeida:2013jfa}, $\Gamma^{\rm SM}(h^0 \to c \bar{c}) =0.118$ MeV~\cite{pdg2014}, $m_b (m_b)^{\msbar}=4.2~\gev, m_c(m_c)^{\msbar}=1.275~\gev$~\cite{PDG2013}, and $\a_s(m_Z)=0.1185$~\cite{a_s@ICHEP2014}. The y-axis counts the number of survived parameter points for each bin of the deviation. It is seen that in the case of $h^0 \to b \bar{b}$ (Fig.~\ref{fig1a}) the detailed variation of the elements $M_U$ and $M_Q$ can increase the effect and the deviation from the SM can go up to $\sim 30\%$ at certain parameter points.
In the case of  $h^0 \to c \bar{c}$ (Fig.~\ref{fig1b}) a large deviation from the SM value due to large values of the product $T^U_{32} M^U_{23}$, discussed at the end of Section 3.2, is in principle possible. 
Since there exists no physical constraint on this product we will only show results with a deviation from the SM up to $\sim \pm 50\%$.
%

%
\begin{table}
\caption{Reference scenario: shown are the basic MSSM parameters 
at $Q = 1$~TeV,
except for $m_{A^0}$ which is the pole mass (i.e.\ the physical mass) of $A^0$, 
with $T_{U33} = 1450$~GeV (corresponding to $\delta^{uRL}_{33} =  0.1$). All other squark 
parameters not shown here are zero. }
\begin{center}
\begin{tabular}{|c|c|c|}
  \hline
 $M_1$ & $M_2$ & $M_3$ \\
 \hline \hline
400 ~\gev  & 800 ~\gev & 2000~\gev \\
  \hline
\end{tabular}
\vskip0.4cm
\begin{tabular}{|c|c|c|}
  \hline
 $\mu$ & $\tan \beta$ & $m_{A^0}$ \\
 \hline \hline
 500~\gev & 30  &  1500 ~\gev \\
  \hline
\end{tabular}
\vskip0.4cm
\begin{tabular}{|c|c|c|c|}
  \hline
   & $\alpha = 1$ & $\alpha= 2$ & $\alpha = 3$ \\
  \hline \hline
   $M_{Q \alpha \alpha}^2$ & $3200^2~\gev^2$ &  $1550^2~\gev^2$  & $1100^2~\gev^2$ \\
   \hline
   $M_{U \alpha \alpha}^2$ & $3200^2~\gev^2$ & $2800^2~\gev^2$ & $2050^2~\gev^2$ \\
   \hline
   $M_{D \alpha \alpha}^2$ & $3200^2~\gev^2$ & $3000^2~\gev^2$ &  $2500^2~\gev^2$  \\
   \hline
\end{tabular}
\vskip0.4cm
\begin{tabular}{|c|c|c|c|}
  \hline
   $\delta^{LL}_{23}$ & $\delta^{uRR}_{23}$  &  $\delta^{uRL}_{23}$ & $\delta^{uLR}_{23}$\\
  \hline \hline
  0  & 0.8 & 0.02  & 0.02  \\
    \hline
\end{tabular}
\end{center}
\label{basicparam}
\end{table}

Based on the results from the scans we have 
chosen a reference scenario with strong $\ti{c}-\ti{t}$ mixing to demonstrate the effects of QFV in both $h^0$ to $b\bar{b}$ and $c\bar{c}$ decays.
The corresponding MSSM parameters at $Q = 1$~TeV are given in Table~\ref{basicparam}. 
%
\begin{table}
\caption{Physical masses in GeV of the particles for the scenario of Table~\ref{basicparam}.}
\begin{center}
\begin{tabular}{|c|c|c|c|c|c|}
  \hline
  $\mnt{1}$ & $\mnt{2}$ & $\mnt{3}$ & $\mnt{4}$ & $\mch{1}$ & $\mch{2}$ \\
  \hline \hline
  $395$ & $507$ & $511$ & $845$ & $501$ & $845$ \\
  \hline
\end{tabular}
\vskip 0.4cm
\begin{tabular}{|c|c|c|c|c|}
  \hline
  $m_{h^0}$ & $m_{H^0}$ & $m_{A^0}$ & $m_{H^+}$ \\
  \hline \hline
  $125$  & $1500$ & $1500$ & $1503$ \\
  \hline
\end{tabular}
\vskip 0.4cm
\begin{tabular}{|c|c|c|c|c|c|c|}
  \hline
  $\msg$ & $\msu{1}$ & $\msu{2}$ & $\msu{3}$ & $\msu{4}$ & $\msu{5}$ & $\msu{6}$ \\
  \hline \hline
  $2103$ & $996$ & $1176$ & $1578$ & $3214$ & $3217$ & $3327$ \\
  \hline
\end{tabular}
\vskip 0.4cm
\begin{tabular}{|c|c|c|c|c|c|}
  \hline
 $\msd{1}$ & $\msd{2}$ & $\msd{3}$ & $\msd{4}$ & $\msd{5}$ & $\msd{6}$ \\
  \hline \hline
  $1128$ & $1579$ & $2515$ & $3012$ & $3211$ & $3218$ \\
  \hline
\end{tabular}
\end{center}
\label{physmasses}
\end{table}
%
\begin{table}
\caption{Flavour decomposition of $\su_i$, $i=1,...,6$ for the scenario of Table~\ref{basicparam}. Shown are the squared coefficients. }
\begin{center}
\begin{tabular}{|c|c|c|c|c|c|c|c|}
  \hline
  & $\su_L$ & $\sca_L$ & $\st_L$ & $\su_R$ & $\sca_R$ & $\st_R$ \\
  \hline \hline
 $\su_1$ & $0$ & $0.002$ & $0.25$ & $0$ & $0.228$ & $0.52$  \\
  \hline 
  $\su_2$  & $0$ & $0$ & $0.749$ & $0$ & $0.086$ & $0.165$ \\
  \hline
  $\su_3$  & $0.051$ & $0.946$ & $0.001$ & $0$ & $0$ & $0$ \\
  \hline
  $\su_4$ & $0.95$ & $0.05$ & $0$ & $0$ & $0$ & $0$  \\
  \hline 
  $\su_5$  & $0$ & $0$ & $0$ & $1$ & $0$ & $0$ \\
  \hline
  $\su_6$  & $0$ & $0$ & $0$ & $0$ & $0.69$ & $0.31$ \\
  \hline
\end{tabular}
\end{center}
\label{flavourdecomp}
\end{table}
%
This scenario satisfies all present experimental and theoretical constraints, see Appendix~\ref{sec:constr}. The resulting physical masses of the particles are shown in Table~\ref{physmasses}. We also show the flavour decomposition of the up-type squarks $\su_i, i=1,...,6$ in Table~\ref{flavourdecomp}. For the calculation of the masses and the mixing, as well as for the low-energy observables, especially 
those in the B meson sector (see Table~\ref{TabConstraints}), we use the public code 
{\tt SPheno} v3.3.3~\cite{SPheno1, SPheno2}. Both the widths $\Gamma(h^0 \to b \bar{b})$ and $\Gamma(h^0 \to c \bar{c})$ are calculated at full one-loop level in the MSSM with QFV  
using the packages {\tt FeynArts}~\cite{Hahn:2000kx} and {\tt FormCalc}~\cite{Hahn:1998yk}. We also use the packages {\tt SSP}~\cite{SSP} and {\tt LoopTools}~\cite{Hahn:1998yk}. For creating the Fortran code for the mass insertion formulas {\tt MassToMI}~\cite{arXiv:1509.05030} was very helpful.
In the following unless specified otherwise we show various parameter dependences of $\Gamma/\Gamma^{\rm SM}-1$ for $\Gamma(h^0 \to b \bar{b})$ and $\Gamma(h^0 \to c \bar{c})$ with all other parameters fixed as in Table~\ref{basicparam}. 
\begin{figure*}
\centering
\subfigure[]{
   { \mbox{\hspace*{-1cm} \resizebox{7.5cm}{!}{\includegraphics{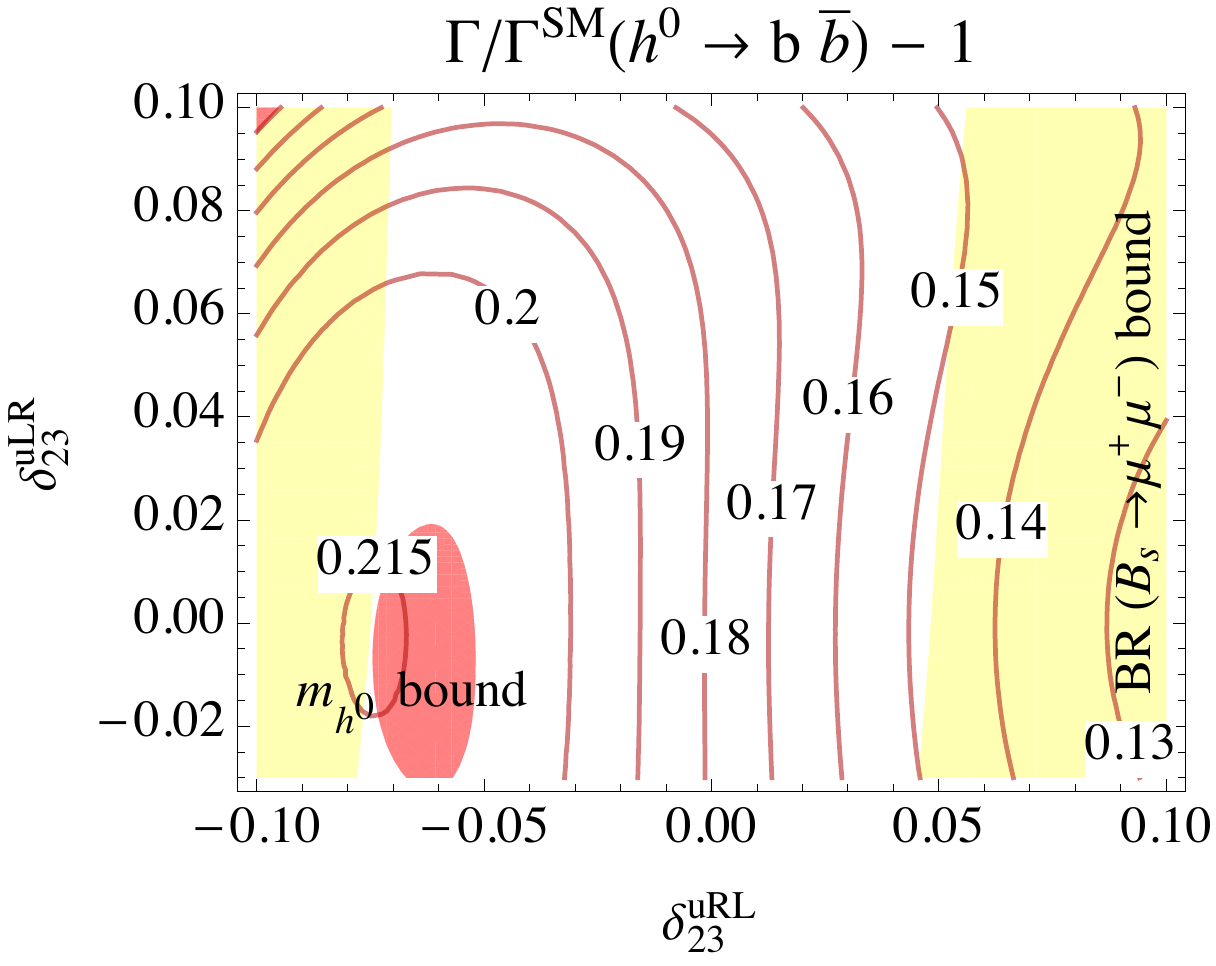}} \hspace*{-0.8cm}}}
   \label{fig2a}
}
 \subfigure[]{
   { \mbox{\hspace*{+0.cm} \resizebox{7.5cm}{!}{\includegraphics{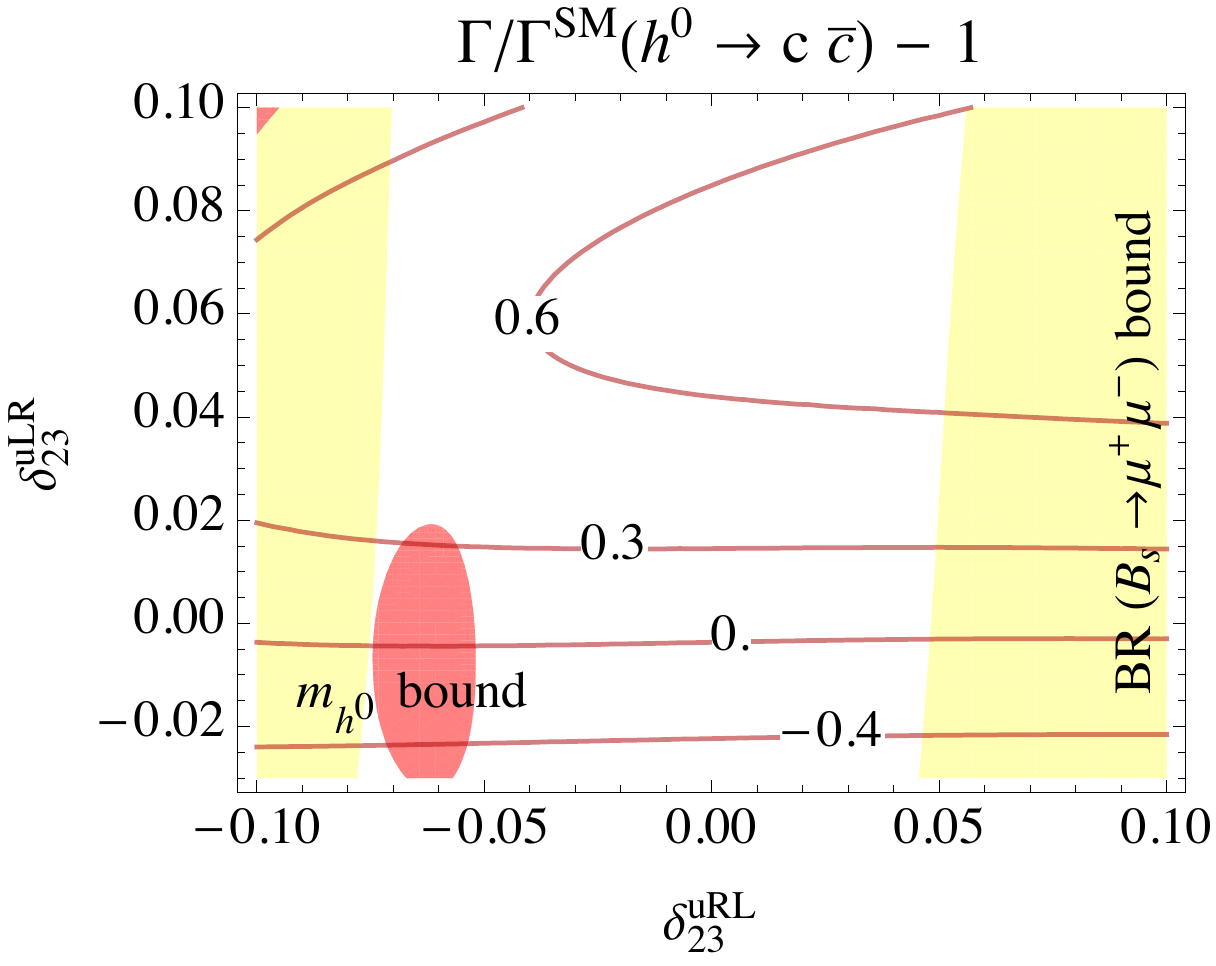}} \hspace*{-1cm}}}
  \label{fig2b}
}
\caption{Contours of the deviation (a)~$\Gamma/\Gamma^{\rm SM}(h^0 \to b \bar{b})-1$ and (b)~$\Gamma/\Gamma^{\rm SM}(h^0 \to c \bar{c})-1$ in the $\durl$-$\dulr$  plane for $\durr = 0.5$ and $\dll=0$.
}
\label{fig2}
\end{figure*}

In Fig.~\ref{fig2} the dependence on the QFV parameters $\durl$ and $\dulr$ is shown. It is seen that in the case of $b\bar{b}$ (Fig.~\ref{fig2a}) the variation due to correlated $\sca_R-\st_L$ and $\sca_L-\st_R$ mixing can vary up to $\sim 6\%$ in the region allowed by the constraints. Comparing Fig.~\ref{fig2a} with Fig.~\ref{fig2b} one can see that there exist regions where both widths considered simultaneously deviate from their SM prediction. Hence $\Gamma(h^0 \to b \bar{b})$ tends to depend more on $\sca_R-\st_L$ mixing, while $\Gamma(h^0 \to c \bar{c})$ depends more on $\sca_L-\st_R$ mixing. 

\begin{figure*}
\centering
\subfigure[]{
   { \mbox{\hspace*{-1cm} \resizebox{7.5cm}{!}{\includegraphics{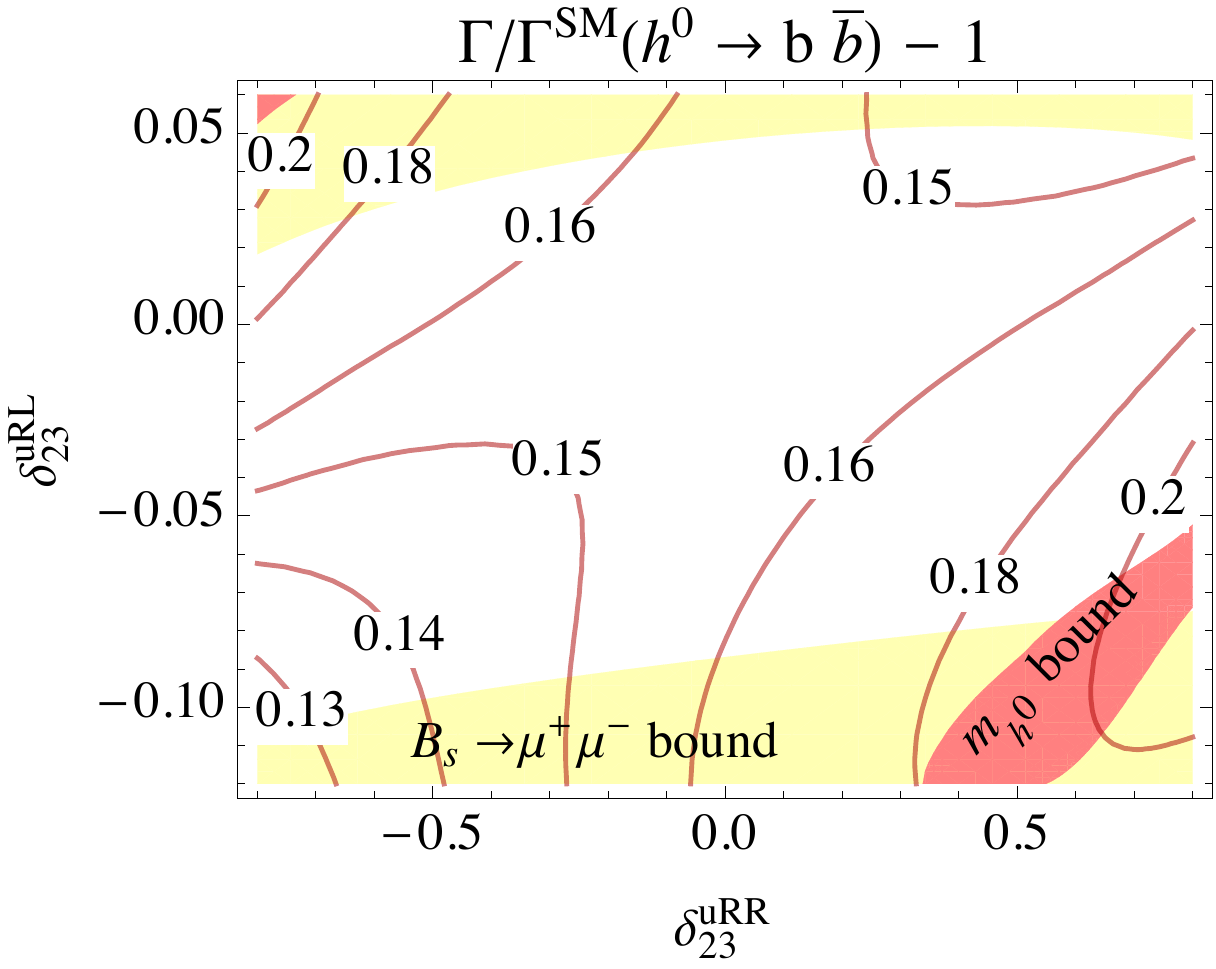}} \hspace*{-0.8cm}}}
   \label{fig3a}
}
 \subfigure[]{
   { \mbox{\hspace*{+0.cm} \resizebox{7.5cm}{!}{\includegraphics{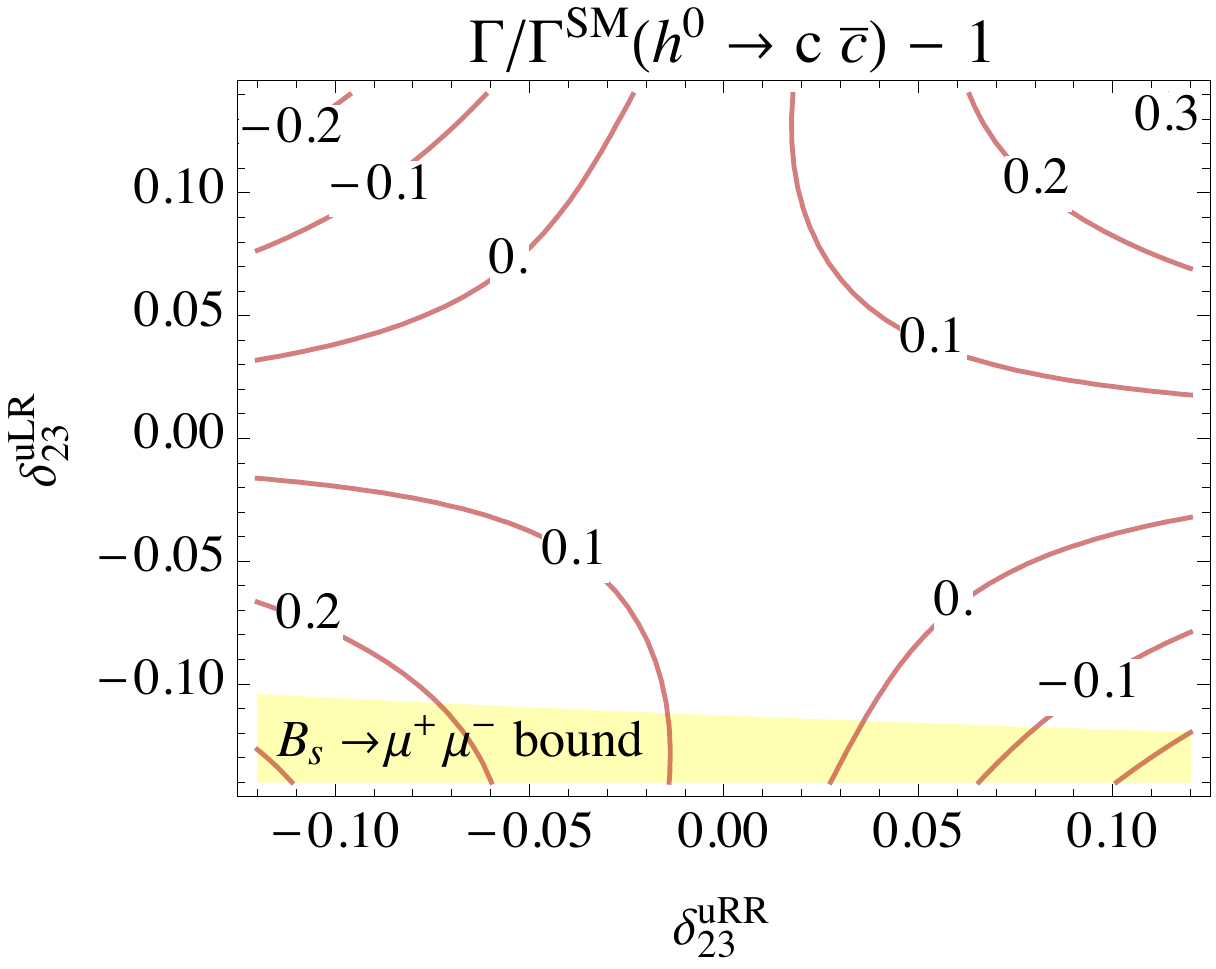}} \hspace*{-1cm}}}
  \label{fig3b}
  }
  \caption{Contours of the deviation (a)~$\Gamma/\Gamma^{\rm SM}(h^0 \to b \bar{b})-1$ in the $\durr$-$\durl$ plane for $\dll = 0$ and $\dulr=0$ and (b)~$\Gamma/\Gamma^{\rm SM}(h^0 \to c \bar{c})-1$ in the $\durr$-$\dulr$ plane for $\dll = 0$ and $\durl=0.02$.
}
\label{fig3}
\end{figure*}

This tendency can also be seen in Fig.~\ref{fig3}. On the left hand side (Fig.~\ref{fig3a}) the dependence of $\Gamma/\Gamma^{\rm SM}(h^0 \to b \bar{b})$ on the QFV parameters $\durr$ and $\durl$ is shown. The variation due to $\sca_R-\st_L$ and $\sca_R-\st_R$ mixing is $\sim 7\%$. In the same scenario, however, the variation of $\Gamma/\Gamma^{\rm SM}(h^0 \to c \bar{c})$ (not shown here) is only $\sim 3\%$. On the right hand side (Fig.~\ref{fig3b}) $\Gamma/\Gamma^{\rm SM}(h^0 \to c \bar{c})$ is shown as a function of $\durr$ and $\dulr$. The variation is large and can go up to $\sim 30\%$, see also~\cite{Bartl:2014bka}. In the same scenario, however, $\Gamma/\Gamma^{\rm SM}(h^0 \to b \bar{b})$ varies only by less than one percent. 
%

\begin{figure*}[h!]
\centering
\subfigure[]{
   { \mbox{\hspace*{-0.5cm} \resizebox{7cm}{!}{\includegraphics{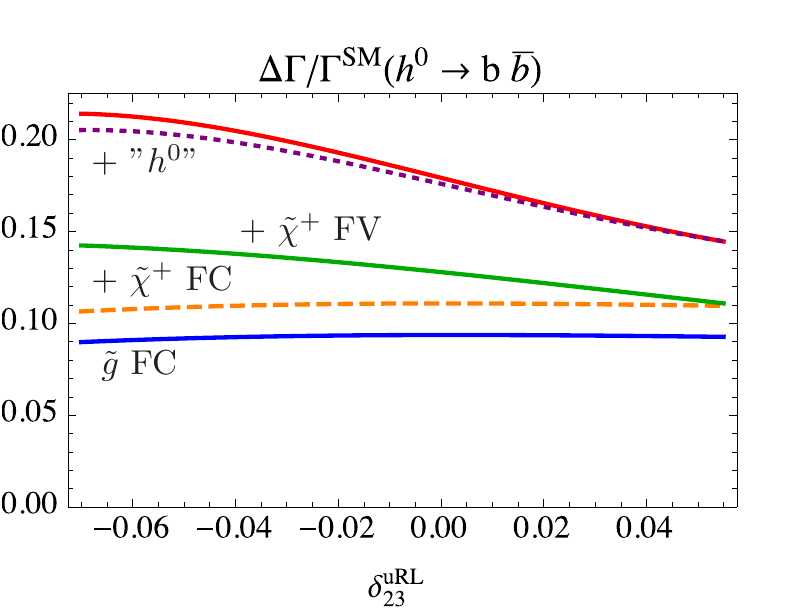}} 
   }}
   \label{MIa}
}
 \subfigure[]{
   { \mbox{\hspace*{+0.cm} \resizebox{7cm}{!}{\includegraphics{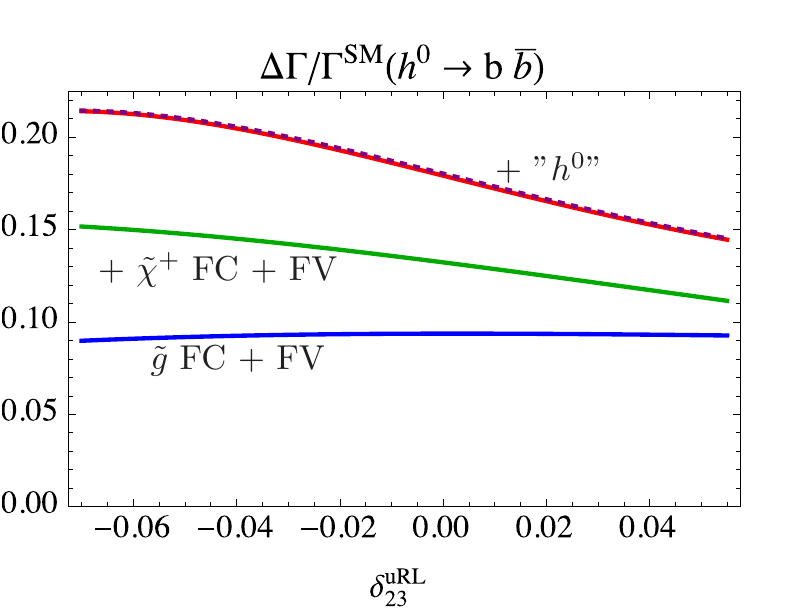}}
   }}
  \label{MIb}
}
\caption{(a) The QFV and QFC gluino and chargino one-loop contributions (added from bottom to top) to $\Gamma/\Gamma^{\rm SM}(h^0 \to b \bar{b})$ computed by using the mass insertion technique, see Section~\ref{Sec:MI}, as a function of $\durl$ for the parameters of Fig.~\ref{fig2a} with $\dulr = 0.02$. (b) The total gluino and chargino one-loop contributions to $\Gamma/\Gamma^{\rm SM}(h^0 \to b \bar{b})$ computed by using the approximate formula~(\ref{MIapprox}) together with eq.~(\ref{dsigLRsg_b}) and the finite part of eq.~(\ref{sigLRbottom}) as a function of $\durl$ for the same parameters as in (a). In both graphs the "$h^0$" contribution denotes $\Gamma^{g, \rm imp}/\Gamma^{\rm SM}(h^0 \to b \bar{b})-1$ and the top curve shows deviation of the full one-loop level width of eq.~(\ref{eqGamma}) from the SM width, $\Gamma/\Gamma^{\rm SM}(h^0 \to b \bar{b})-1$, with no approximation.
}
\label{figMI}
\end{figure*}

In Section~\ref{SecMI:gluino2ccb}, in agreement with our results in Ref.~\cite{Bartl:2014bka}, we have shown that in the case of $c\bar{c}$ the deviation from the SM is entirely due to QFV. However, it is known that in the MSSM $\Gamma(h^0 \to b\bar{b})$ can differ considerably from the SM due to QFC contributions~\cite{Endo:2015oia}.
In Fig.~\ref{MIa} the individual contributions to $\Gamma/\Gamma^{\rm SM}(h^0 \to b\bar{b})$ 
(i.e. QFC gluino one-loop, QFC and QFV chargino one-loop contributions computed 
in the mass insertion approximation) are shown as a function of $\durl$ for the parameters 
of Fig.~\ref{fig2a} with $\durl$=0.02. 
The QFC/QFV gluino and chargino one-loop contributions 
in the mass insertion approximation are given in Section~\ref{Sec:MI}. The $"h^0"$ contribution denotes 
$\Gamma^{g, \rm impr}/\Gamma^{\rm SM}(h^0 \to b\bar{b}) - 1$ which depends on $m_{h^0}$ and the angle $\a$ 
and hence depends on both the QFC and QFV parameters. Note that $m_{h^0}$ as well as $\sin \a$ 
already appear in the kinematics factor at tree level, see eq.~(\ref{decaywidttree}). The top curve shows 
the deviation of the full one-loop level width of eq.~(\ref{eqGamma}) from the SM width, 
$\Gamma/\Gamma^{\rm SM}(h^0 \to b\bar{b}) - 1$, with no approximation.
It is seen that the main one-loop contributions to $\Gamma(h^0 \to b\bar{b})$ come 
from QFC gluino and QFC chargino exchange. Nevertheless, there exists a region 
for large and negative $\durl$ where the QFV component can be comparable with 
the QFC component. The QFV component is mainly due to chargino exchange which 
involves mixing in the $\su$ sector. On the other hand, in the $b \bar{b}$ case the 
gluino exchange, which plays a major role in the $c \bar{c}$ case, involves $\sd$ quarks 
whose QFV mixing effect is strongly suppressed, and hence the QFV component of 
the gluino exchange contribution is very small. Therefore, it is not shown in this 
figure. It is also interesting that the $"h^0"$ contribution depends significantly 
on the QFV parameter $\durl$. 
After all, the variation of $\Gamma/\Gamma^{\rm SM}(h^0 \to b\bar{b})$ in the shown QFV 
parameter range, which can be taken as QFV effect, can be 
as large as $\sim 7\%.$

Fig.\ref{MIb} demonstrates the quality of our approximated result~(\ref{MIapprox}).
By comparing numerically the different MI orders we realized that the MI formulas converge
fast for $\tilde g$~FC and $\tilde \chi^+$~FC, but not for $\tilde \chi^+$~FV.
This can be seen by comparing Fig.~\ref{MIa} with Fig.~\ref{MIb}.
Thus, the difference between the dotted curve and the
upper curve in Fig.~\ref{MIa} is mainly due to the relatively slow MI convergence of the $\tilde \chi^+$~FV contribution.

Although the decay $h^0 \to b \bar{b}$ is dominant, the measurement of its branching ratio and width at the LHC will be a big challenge. At LHC one always measures $\sigma(pp\to h^0 X){\rm B}(h^0 \to b \bar{b})$.
The largest Higgs boson production cross section is due to gluon gluon fusion. However, due to the huge QCD background it will be difficult to isolate the $h^0 \to b \bar{b}$ mode.
The other production modes (vector boson fusion, Higgs radiation from $W^\pm Z$, and associated $t \bar{t}h^0$ production) have smaller cross sections, but may have less background. In any case, high luminosity at LHC would be needed~\cite{CMS:2013xfa}. A model independent and precise measurement of B($h^0 \to b \bar{b}$) and $\Gamma(h^0 \to b \bar{b})$ would be possible at a $e^+ e^-$ linear collider such as ILC~\cite{Barklow:2015tja}.

\section{Conclusions}
\label{sec:concl}

In analogy to our previous paper~\cite{Bartl:2014bka}, we have calculated the decay width of $h^0 \to b\bar{b}$ in the MSSM with quark flavour violation at full one-loop level. We have studied the effects of $\sca-\st$ mixing, taking into account all constraints on the QFV parameters from B-meson data. We have discussed in detail both the decays $h^0 \to c \bar{c}$ and $h^0 \to b \bar{b}$ within the perturbative mass insertion technique applying the Flavour Expansion Theorem~\cite{Dedes:2015twa}. 
There are cases, where the charm self-energy and consequently the correction to the width $\Gamma(h^0 \to c \bar{c})$ can become unacceptably large. This is due to the product $M^U_{23} T^U_{32}$, for which there exists no bound.
%
In general, the deviation of $\Gamma(h^0 \to b \bar{b})$ from the SM can be large (up to 30\%), mainly coming from the QFC part of the MSSM. The QFV contribution due to $\sca_{L,R}-\st_{L,R}$ mixing and chargino exchange is smaller but can nevertheless reach $\sim 7\%$ at certain parameter points. The QFV part due to gluino exchange, which is due to $\ss_{L,R}-\sb_{L,R}$ mixing, is very small.

\begin{appendix}

\section{Interaction Lagrangian}
\label{sec:lag}

\begin{itemize}

\item In the MSSM the interaction of the lightest neutral Higgs boson, $h^0$, with two bottom quarks is given by
\be
{\cal L}_{h^0 b \bar{b}}= s_1^b h^0  \bar{b} b\,,
\label{treelag}
\ee
with the tree-level coupling $s_1^b$ given by eq.~(\ref{treecoup}).

\item  In the super-CKM basis, the interaction of the 
lightest neutral Higgs boson, $h^0$, with two down-type squarks is given by
\be
{\cal L}_{h^0\sd_i \sd_j}=G_{ij1}^{\sd} h^0 \sd_j^* \sd_i, \, ~ i,j=1,...,6.
\ee
The coupling $G_{ij1}^{\sd}$ reads
\begin{eqnarray}
&&  G_{ij1}^{\sd}=\frac{g} {2 m_W}\,
	\bigg[ -m_W^2 \sin(\alpha+\beta) \Big[
      (1+\tfrac13 \tan^2\thw) \nonumber \\
    &&  \times (U^{\sd})_{jk} (U^{\sd *})_{ik} + \tfrac23 \tan^2\thw (U^{\sd})_{j\,(k+3)} 
    (U^{\sd *})_{i\,(k+3)}
      \Big] \nonumber \\[0.2cm]
    & & +\ 2 \dfrac{\sin\a}{\cos\b} \Big[
      (U^{\sd})_{jk}\ m^2_{d,k} (U^{\sd *})_{ik} 
      + (U^{\sd})_{j\,(k+3)} m^2_{d,k} (U^{\sd *})_{i\,(k+3)}
      \Big] \nonumber \\[0.2cm]
    & & +\ \dfrac{\cos\a}{\cos\b} \Big[
      \mu^* (U^{\sd})_{j\,(k+3)} m_{d,k} (U^{\sd *})_{ik} 
      + \mu (U^{\sd})_{jk} m_{d,k} (U^{\sd *})_{i\,(k+3)}
      \Big] \nonumber \\[0.2cm]
    & & +\ \dfrac{\sin\a}{\cos\b}\, \dfrac{v_1}{\sqrt2} \Big[
      (U^{\sd})_{j(k+3)}\ (T_D)_{kl}\ (U^{\sd *})_{il} 
      + (U^{\sd})_{jk}\ (T_D^*)_{lk}\ (U^{\sd *})_{i\,(l+3)}
    \Big] \bigg] \, ,
    \label{coupldidjh}
\end{eqnarray}
where the sum over $k,l=1,2,3$ is understood.
Here $U^{\sd}$ is the mixing matrix of the down-type squarks
\begin{eqnarray}
&&\sd_{kL} = (U^{\sd \dagger})_{ki} \sd_i, \, \nonumber \\ 
&&\sd_{kR} = (U^{\sd \dagger})_{(k + 3)\,i} \sd_i, \, ~ k=1,2,3,~ i=1,...,6.
\end{eqnarray}
Note that $(T_D)_{kl}$ in eq.~(\ref{coupldidjh}) are given in the SUSY Les Houche Accord notation\cite{Skands:2003cj}.

\item  The interaction of gluino, down-type squark and a bottom quark is given by
\bea
{\cal L}_{\sg \sd_j b}&=& -\sqrt{2} g_s T_{rl}^{\a}\bigg[\bar{b}^r
(U^{\sd *}_{j3} e^{i\frac{\phi_3}{2}}P_R-U^{\sd *}_{j6}e^{-i\frac{\phi_3}{2}} P_L) \sg^{\a} \sd_j^l \nn
&&+\bar{\sg}^{\a}(U^{\sd}_{j3}e^{-i\frac{\phi_3}{2}}P_L-U^{\sd}_{j6}e^{i\frac{\phi_3}{2}}P_R) b^l \sd_j^{*, r}
\bigg]\,,
\eea
where $T^\a$ are the SU(3) colour group generators and summation over $r,l=1,2,3$ and over $\a=1,...,8$ is understood. In our case the parameter $M_3=\msg e^{i \phi_3}$ is taken as real, $\phi_3=0$.

\item  The interaction of chargino, up-type squark and a bottom quark is given by
\bea
{\cal L}_{\chp_m b \su_i}= \bar{b} \left(k_{im}^{\su}P_L + l_{im}^{\su}P_R \right)  \ti{\x}_m^{+ *} \su_i + \overline{ \ti{\x}_m^{+ *}} \left(k_{im}^{\su*}P_R + l_{im}^{\su*}P_L \right) b~\su_i^*\,,
\eea
where the couplings $k_{im}^{\su}$ and $ l_{im}^{\su}$ are given by
\bea
&& k_{im}^{\su} = h_b U^*_{m 2} U^{\su *}_{i3} \, \nonumber \\ 
&& l_{im}^{\su} = -g V_{m 1} U^{\su *}_{i3} + h_t V_{m 2} U^{\su *}_{i6}\, 
\eea
\begin{table*}
\footnotesize{
\caption{
Constraints on the MSSM parameters from the B-physics experiments
relevant mainly for the mixing between the 2nd and the 3rd generations of 
squarks and from the data on the $h^0$ mass. The last column shows the constraints 
at $95 \%$ CL obtained by combining the experimental error quadratically
with the theoretical uncertainty, except for $m_{h^0}$, see Ref.~\cite{Bartl:2014bka}.
}
\begin{center}
\begin{tabular}{|c|c|c|c|}
    \hline
    Observable & Exp.\ data & Theor.\ uncertainty & \ Constr.\ (95$\%$CL) \\
    \hline\hline
    &&&\\
    $\Delta M_{B_s}$ [ps$^{-1}$] & $17.757 \pm 0.021$ (68$\%$ CL)~\cite{DeltaMBs_HFAG2014} 
    & $\pm 3.3$ (95$\%$ CL)~\cite{DeltaMBs_Carena2006, Ball_2006} &
    $17.757 \pm 3.30$\\

    $10^4\times$B($b \to s \gamma)$ & $3.41 \pm 0.155$ (68$\%$ CL)~\cite{Trabelsi_EPS-HEP2015} 
    & $\pm 0.23$ (68$\%$ CL)~\cite{Misiak_2015} &  $3.41\pm 0.54$\\

    $10^6\times$B($b \to s~l^+ l^-$)& $1.60 ~ ^{+0.48}_{-0.45}$ (68$\%$ CL)~\cite{bsll_BABAR_2014}
    & $\pm 0.11$ (68$\%$ CL)~\cite{Huber_2008} & $1.60 ~ ^{+0.97}_{-0.91}$\\
    $(l=e~{\rm or}~\mu)$ &&&\\

    $10^9\times$B($B_s\to \mu^+\mu^-$) & $2.8~^{+0.7}_{-0.6}$ (68$\%$CL)~\cite{Bsmumu_LHCb_CMS}
    & $\pm0.23$  (68$\%$ CL)~\cite{Bsmumu_SM_Bobeth_2014} 
    & $2.80~^{+1.44}_{-1.26}$ \\

    $10^4\times$B($B^+ \to \tau^+ \nu $) & $1.14 \pm 0.27$ (68$\%$CL)
    ~\cite{Trabelsi_EPS-HEP2015, Hamer_EPS-HEP2015}
    &$\pm0.29$  (68$\%$ CL)~\cite{Btotaunu_LP2013} & $1.14 \pm 0.78$\\

    $ m_{h^0}$ [GeV] & $125.09 \pm 0.24~(68\%~ \rm{CL})$ \cite{Higgs_mass_ATLAS_CMS}
    & $\pm 3$~\cite{Higgs_mass_Heinemeyer} & $125.09 \pm 3.48$ \\

&&&\\
    \hline
\end{tabular}
\end{center}
\label{TabConstraints}}
\end{table*}
%
$U$ and $V$ are unitary matrices that diagonalise the charging mass matrix 
$U^* X V^\dagger = {\rm diag}(m_{ \ti{\x}_1^{\pm}}, m_{ \ti{\x}_2^{\pm}})$ and $h_{t,b}$ are the top and bottom Yukawa couplings $h_{t(b)}=\tfrac{g m_{t (b)}}{\sqrt{2} m_W \sin \b (\cos \b)}$.

The interaction Lagrangian for the $h^0 \to c \bar{c}$ case is given in Ref.~\cite{Bartl:2014bka}.

\end{itemize}

\section{Theoretical and experimental constraints}
\label{sec:constr}

The experimental and theoretical constraints taken into 
account in the present note are discussed in detail in Ref.~\cite{Bartl:2014bka}. Here we only list the updated constraints from 
B-physics and those on the Higgs boson mass in 
Table~\ref{TabConstraints}.

\end{appendix}

%
\section*{Acknowledgments}

This work is supported by the "Fonds zur F\"orderung der
wissenschaftlichen Forschung (FWF)" of Austria, project No. P26338-N27.

%
%


\begin{thebibliography}{99}

\bibitem{Aad:2012tfa}
  G.~Aad {\it et al.} [ATLAS Collaboration],
  Phys. Lett. B  716 (2012) 1
  [arXiv:1207.7214 [hep-ex]].
  
\bibitem{Chatrchyan:2012xdj}
  \mbox{S.~Chatrchyan {\it et al.} [CMS Collaboration],
  Phys. Lett. B 716 (2012) 30} 
  [arXiv:1207.7235 [hep-ex]].
  
\bibitem{pdg2014}
K~.A~. Olive {\it et al.}~(Particle Data Group), Chin. Phys. C, 38, 090001 (2014) and 2015 update.

\bibitem{Bartl:2014bka}
  A.~Bartl, H.~Eberl, E.~Ginina, K.~Hidaka and W.~Majerotto,
  Phys. Rev. D 91 (2015) no.1,  015007
  [arXiv:1411.2840 [hep-ph]].
  
\bibitem{Eberl:2014dla}
  H.~Eberl, A.~Bartl, E.~Ginina, K.~Hidaka and W.~Majerotto,
  [arXiv:1412.5392 [hep-ph]].
  
\bibitem{Hidaka:2015zqa}
  K.~Hidaka, A.~Bartl, H.~Eberl, E.~Ginina and W.~Majerotto,
  [arXiv:1504.07792 [hep-ph]].
  
\bibitem{Ginina:2015doa}
  E.~Ginina, H.~Eberl, W.~Majerotto, A.~Bartl and K.~Hidaka,
  PoS EPS-HEP2015 (2015) 146
  [arXiv:1510.03714 [hep-ph]].
  
\bibitem{Hidaka:2015ewt}
  K.~Hidaka, A.~Bartl, H.~Eberl, E.~Ginina and W.~Majerotto,
  PoS EPS-HEP2015 (2015) 131
  [arXiv:1511.01977 [hep-ph]].
  
\bibitem{Dedes:2015twa}
  A.~Dedes, M.~Paraskevas, J.~Rosiek, K.~Suxho and K.~Tamvakis,
  JHEP 1506 (2015) 151
  [arXiv:1504.00960 [hep-ph]].

\bibitem{Allanach:2008qq}
  B.~C.~Allanach {\it et al.},
  Comput. Phys. Commun. 180 (2009) 8
  [arXiv:0801.0045 [hep-ph]].
 
\bibitem{Gabbiani:1996hi}
  F.~Gabbiani, E.~Gabrielli, A.~Masiero and L.~Silvestrini,
  Nucl. Phys. B 477 (1996) 321
  [hep-ph/9604387].
  
\bibitem{Dedes:2014asa}
  A.~Dedes, M.~Paraskevas, J.~Rosiek, K.~Suxho and K.~Tamvakis,
  JHEP {\bf 1411} (2014) 137
  [arXiv:1409.6546 [hep-ph]].
  
  \bibitem{G&H}
J.~F.~Gunion, H.~E.~Haber, Nucl. Phys. B272 (1986) 1.


\bibitem{Carena:1999py}
  M.~Carena, D.~Garcia, U.~Nierste and C.~E.~M.~Wagner,
  Nucl. Phys. B 577 (2000) 88
  [hep-ph/9912516].
 
\bibitem{arXiv:1509.05030}
J. Rosiek, 
Comput. Phys. Commun. 201 (2016) 144
[arXiv:1509.05030].

\bibitem{PV}
G. Passarino, M.J.G. Veltman, Nucl. Phys. B {\bf 160} (1979) 151.

\bibitem{Brignole:2015kva}
  A.~Brignole,
  Nucl. Phys. B 898 (2015) 644
  [arXiv:1504.03273 [hep-ph]].
  
\bibitem{Crivellin:2010er}
  A.~Crivellin,
  Phys. Rev. D 83 (2011) 056001
  [arXiv:1012.4840 [hep-ph]].
  
\bibitem{Crivellin:2012zz}
  A.~Crivellin and C.~Greub,
  Phys. Rev. D 87 (2013) 015013
  [arXiv:1210.7453 [hep-ph]].


\bibitem{DeCausmaecker:2015yca}
  K.~De Causmaecker, B.~Fuks, B.~Herrmann, F.~Mahmoudi, B.~O'Leary, W.~Porod, S.~Sekmen and N.~Strobbe,
  JHEP 1511 (2015) 125
  [arXiv:1509.05414 [hep-ph]].

\bibitem{Almeida:2013jfa}
  L.~G.~Almeida, S.~J.~Lee, S.~Pokorski and J.~D.~Wells,
  Phys. Rev. D 89 (2014) no.3,  033006
  [arXiv:1311.6721 [hep-ph]].
  
  \bibitem{PDG2013}
  J. Beringer et al. (Particle Data Group), Phys.\ Rev.\ D {\bf 86} (2012) 010001.
  
\bibitem{a_s@ICHEP2014}                                   
   C. ~Roda, plenary talk at 37th International Conference on
   High Energy Physics, Valencia, Spain, 2-9 July 2014.
 
%
\bibitem{SPheno1}  
W.~Porod, Comput. Phys. Commun. 153 (2003) 275 [hep-ph/0301101].  

\bibitem{SPheno2}  
W.~Porod and F.~Staub, Comput. Phys. Commun. 183 (2012) 2458 [arXiv:1104.1573
[hep-ph]].

\bibitem{Hahn:2000kx}
  T.~Hahn,
  Comput. Phys. Commun.140 (2001) 418
  [hep-ph/0012260].
  
\bibitem{Hahn:1998yk}
  T.~Hahn and M.~Perez-Victoria,
  Comput. Phys. Commun. 118 (1999) 153
  [hep-ph/9807565].
  
\bibitem{SSP}  
F.~Staub, T.~Ohl, W.~Porod, C.~Speckner, Computer Physics Communications 183 (2012) 2165.

\bibitem{Endo:2015oia}
  M.~Endo, T.~Moroi and M.~M.~Nojiri,
  JHEP 1504 (2015) 176
  [arXiv:1502.03959 [hep-ph]].
 
 
\bibitem{CMS:2013xfa}
  [CMS Collaboration],
  [arXiv:1307.7135 [hep-ex]].
  
\bibitem{Barklow:2015tja}
  T.~Barklow, J.~Brau, K.~Fujii, J.~Gao, J.~List, N.~Walker and K.~Yokoya,
  [arXiv:1506.07830 [hep-ex]].
 
 
 %
\bibitem{Skands:2003cj}
  P.~Z.~Skands, B.~C.~Allanach, H.~Baer, C.~Balazs, G.~Belanger, F.~Boudjema, A.~Djouadi and R.~Godbole {\it et al.},
  JHEP 0407 (2004) 036
  [hep-ph/0311123].
  
\bibitem{DeltaMBs_HFAG2014}
  Y.~Amhis {\it et al.} [Heavy Flavor Averaging Group (HFAG) Collaboration],
  [arXiv:1412.7515 [hep-ex]].
%

\bibitem{DeltaMBs_Carena2006}
  M.~S.~Carena  {\it et al.}, 
  Phys. Rev. D 74 (2006) 015009 [hep-ph/0603106].

%
\bibitem{Ball_2006}
  P.~Ball and R.~Fleischer,
  Eur. Phys. J. C 48 (2006) 413 [hep-ph/0604249].
  
 \bibitem{Trabelsi_EPS-HEP2015} 
  K. ~Trabelsi, plenary talk at European Physical Society Conference on High Energy 
  Physics 2015 (EPS-HEP2015), Vienna, 22 - 29 July 2015. 

 \bibitem{Misiak_2015}
   M. Misiak et al., Phys. Rev. Lett. 114 (2015) 221801 [arXiv:1503.01789[hep-ph]].

%
\bibitem{bsll_BABAR_2014}
  J.P. ~Lees {\it et al.}  [BABAR Collaboration],
  Phys. Rev. Lett. 112 (2014) 211802 [arXiv:1312.5364 [hep-ex]].

%
\bibitem{Huber_2008}
  T.~Huber, T.~Hurth and E.~Lunghi,
  Nucl. Phys. B 802 (2008) 40 [arXiv:0712.3009 [hep-ph]].

%
 \bibitem{Bsmumu_LHCb_CMS}
   V. Khachatryan et al. [CMS and LHCb Collaborations],
   Nature 522 (2015) 68 [arXiv:1411.4413[hep-ex]].
 
%
\bibitem{Bsmumu_SM_Bobeth_2014}
  C. ~Bobeth {\it et al.},
  Phys. Rev. Lett. 112 (2014) 101801 [arXiv:1311.0903 [hep-ph]].
  
   \bibitem{Hamer_EPS-HEP2015} 
  P. ~Hamer, talk at European Physical Society Conference on High Energy 
  Physics 2015 (EPS-HEP2015), Vienna, 22 - 29 July 2015. 
  
%
\bibitem{Btotaunu_LP2013}
 J. M. Roney, "Results from the B-Factories", talk at 26th International Symposium on Lepton Photon
 Interactions at High Energies, San Francisco, USA, 24-29 June 2013.

%
\bibitem{Higgs_mass_ATLAS_CMS} 
  ATLAS and CMS collaborations, Phys. Rev. Lett. 114 (2015) 191803, 
  [arXiv:1503.07589[hep-ex]].

%
\bibitem{Higgs_mass_Heinemeyer}
  S. Borowka, T. Hahn, S. Heinemeyer, G. Heinrich and W. Hollik, 
  Eur. Phys. J. C75 (2015) 424 [arXiv:1505.03133 [hep-ph]].



\end{thebibliography}
\end{document}